\documentclass{article}

\usepackage{arxiv}

\usepackage[utf8]{inputenc} 
\usepackage[T1]{fontenc}    
\usepackage{hyperref}       
\usepackage{url}            
\usepackage{booktabs}       
\usepackage{amsfonts}       
\usepackage{nicefrac}       
\usepackage{microtype}      
\usepackage{lipsum}
\usepackage{graphicx}
\usepackage{natbib}
\usepackage{tikz}
\usepackage{bbm}
\RequirePackage{algorithm}%
\RequirePackage{algorithmicx}%
\RequirePackage{algpseudocode}%
\RequirePackage{listings}%
\usetikzlibrary{shapes.geometric, arrows, calc}
\graphicspath{ {./images/} }

\tikzstyle{block} = [rectangle, draw, rounded corners, text centered, text width=11.5em, minimum height=3em]
\tikzstyle{arrow} = [thick, ->, >=stealth]
\newcommand{\ind}{\perp\!\!\!\!\perp}

\usepackage{amsmath}
\usepackage{amsthm}
\theoremstyle{plain}
\newtheorem{assumption}{Assumption}

\title{Gender disparities in rehospitalisations after coronary artery bypass grafting: evidence from a sparse functional causal mediation analysis of the MIMIC-IV data}

\author{
 Henan Xu \\
  Department of Statistics and Actuarial Science\\
  University of Waterloo\\
  Waterloo, ON N2L 3G1 \\
  \texttt{henan.xu@uwaterloo.ca} \\
   \And
 Yeying Zhu \\
  Department of Statistics and Actuarial Science\\
  University of Waterloo\\
  Waterloo, ON N2L 3G1  \\
  \texttt{yeying.zhu@uwaterloo.ca} \\
  \And
 Donna L.Coffman \\
  Department of Psychology\\
  University of South Carolina\\
  Columbia, SC 29208 \\
  \texttt{DCOFFMAN@mailbox.sc.edu} \\
}

\begin{document}
\maketitle
\begin{abstract}
Hospital readmissions following coronary artery bypass grafting (CABG) not only impose a substantial cost burden on healthcare systems but also serve as a potential indicator of the quality of medical care. Previous studies of gender effects on complications after CABG surgery have consistently revealed that women tend to suffer worse outcomes. To better understand the causal pathway from gender to the number of rehospitalisations, we study the postoperative central venous pressure (CVP), recorded over the first 24 hours of patients' intensive care unit (ICU) stay after the CABG surgery, as sparse observations of a functional mediator. Confronted with time-varying CVP measurements and zero-inflated rehospitalisation counts within 60 days following discharge, we propose a parameter-simulating quasi-Bayesian Monte Carlo approximation method that accommodates a sparse functional mediator and a zero-inflated count outcome for causal mediation analysis. We find a causal relationship between the female gender and increased rehospitalisation counts after CABG, and that time-varying central venous pressure mediates this causal effect.
\end{abstract}


\section{Introduction} \label{sec1}
Hospital readmissions are a primary target to improve for both policymakers and clinical administrators, as they are expensive to healthcare systems and are an implication of the quality of care \citep{joynt2011thirty, tsai2013variation, iribarne2014readmissions}. The total number of rehospitalisations within 30 days in 2020 for patients in the United States is over 3.4 million with an average readmission rate of 14\%, imposing a combined annual cost of over \$59 billion on all payers \citep{jiang2024clinical}. Moreover, as suggested by \citet{tsai2013variation}, the rate of postoperative readmissions is significantly lower in hospitals with high surgical volume and low rates of surgical mortality. The fact that the reduction in readmissions is related to other quality measures of hospital care further reinforces it as a priority for clinical practitioners.

Coronary artery bypass graft (CABG) surgery, a procedure that restores blood flow to the heart by redirecting blood around blocked or narrowed coronary arteries using grafts, accounts for more than half of all adult cardiac surgical procedures \citep{2024HDSS}. The prevalence of CABG, along with its hefty \$13,499 mean cost of readmission \citep{shah2019incidence}, make CABG a key focus for reducing healthcare expenses.

Female gender has been consistently reported by existing literature to be a significant risk factor for various postoperative complications following CABG. This raises another concern in addition to healthcare expenditures: the problem of health equity. \citet{enumah2020persistent} and \citet{wagner2024sex} find that women tend to suffer more postoperative comorbidities, as well as higher odds of death. Readmissions, as a potential measure of medical care received during the primary hospital stay, further highlight the issue of gender disparities. For over two decades, analyses of readmissions following CABG have persistently revealed that female patients are more likely to be rehospitalised within a short period after discharge \citep{zitser1999prediction, stewart2000predictors, steuer2002hospital, hannan2003predictors, feng2018coronary, shah2019incidence, shawon2021patient}. Attempts to elucidate the underlying mechanisms of such gender differences are of great interest to health research since such understandings prevent adverse decisions based on purely associational conclusions. For example, referrals of women for CABG could be delayed according to such findings \citep{aldea1999effect}, which could further undermine the well-being of female patients in need. While more recent studies have begun investigating the specific reasons behind the worse outcomes for women, the area remains largely understudied. Some of the hypotheses made from discoveries by \citet{wagner2024sex} include higher comorbidity burdens, underdiagnosis, undertreatment, delays in treatment, as well as biological differences for female ischemic heart disease patients.

While gender is often the exposure of interest in studies of health disparities, its use as a treatment variable in causal inference frameworks has been subject to fierce conceptual debate. Some scholars argue that causal reasoning requires well-defined hypothetical interventions, which are ill-suited to immutable traits such as gender or race \citep{holland1986statistics, hernan2004definition, hernan2008does, kaufman2008epidemiologic, didden2025targeting}. These critiques question whether meaningful counterfactuals can be defined when the exposure itself is not manipulable in practice or even in principle. However, a growing body of work argues that causal reasoning remains valid in such settings if the focus is on policy-relevant contrasts or on decomposing disparities via modifiable mediators. Notably, \citet{vanderweele2012natural, vanderweele2014causal, glymour2014commentary} argue that causal estimands involving non-manipulable exposures can be interpreted as comparisons across hypothetical populations governed by different social or biological assignments, provided careful framing and appropriate interpretation. \citet{pearl2018does} further asserts that causality does not require manipulability as long as the structural dependencies are clear. Our rationale for conceptualising gender as a treatment in the context of conditioning on the CABG population, and the broader debate surrounding non-manipulable treatments, are addressed in greater detail in Section \ref{sec5}.

Leveraging the vast amount of data in the Medical Information Mart for Intensive Care IV (MIMIC-IV) database \citep{johnson2023mimiciv}, we aim to investigate gender disparities in hospital readmission counts following CABG surgery, with a focus on the mediating role of time-varying central venous pressure during the first 24 hours of post-surgical ICU stay through the lens of causal mediation analysis. Central venous pressure (CVP) is a hemodynamic parameter that measures the pressure in the large veins near the heart, indicating blood volume and heart function. Although CVP is routinely measured and documented, its potential as a prognostic indicator of postoperative outcomes in patients following CABG has received insufficient attention \citep{williams2014central}. In fact, \citet{williams2014central} shows that CVP measured 6 hours post-surgery is associated with operative mortality and renal failure after CABG. Our research question in this paper is two-fold. Firstly, we would like to investigate the causal effect of gender on the number of hospital readmissions within 60 days of discharge from the primary hospital stay among patients who underwent CABG in the MIMIC-IV database. Secondly, we intend to examine whether the time-varying CVP measured over the first 24 hours of post-surgical ICU stay mediates the causal pathway from gender to rehospitalisations. Consequently, we arrive at a functional causal mediation analysis with the 60-day readmission count as a zero-inflated count outcome, with gender being a binary treatment and time-varying CVP during the first 24 hours of postoperative ICU stay being an irregularly observed functional mediator.

Applying functional data analysis (FDA) techniques to causal mediation analysis with time-varying mediators commenced fairly recently with the work of \citet{lindquist12}. \citet{lindquist12} considers a framework where the mediator is a continuous function of time observed on a dense grid, the treatment assignment is a binary univariate scalar, and the outcome is a continuous univariate scalar. Subsequently, linear functional structural equation models (lfSEM), as a functional analogue to a linear SEM, are proposed to assess mediation. \citet{coffman2023causal} proposes a similar framework to study the effect of varenicline on smoking cessation, mediated by time-varying craving to smoke. A causal mediation framework is introduced to account for a binary treatment comparing varenicline to nicotine replacement therapy, with time-varying smoking cravings, collected via EMA prompts, as a functional mediator, and a distal binary abstinence outcome. However, subjects with a significant number of missed prompts are discarded to facilitate the use of traditional FDA techniques. The same topic is also explored by \citet{cai2022estimation}, accommodating both a functional mediator and a functional outcome by specifying an SEM with functional response regressions. \citet{zeng2021causal} further advances functional causal mediation analysis by addressing infrequent and irregular longitudinal data, employing functional principal component analysis (FPCA) and a Bayesian approach to estimate direct and indirect effects, particularly in settings where the mediator and outcome are observed on irregular time grids such as the animal behaviour data studied. \citet{zhaoandluo} develops a framework that combines mediation analysis with Granger causality to capture spatio-temporal dependencies in fMRI time series, providing insights into brain mechanisms. \citet{zhao2018functional, zhaou2024unpublished} extend the setting further to incorporate dense functional data in all of the treatment, mediator and outcome. Specifically, they develop two models: the concurrent mediation model, which assumes point-wise relationships at each time point, and the historical mediation model, which accounts for cumulative effects over time. These models allow for the estimation of time-varying direct and indirect effects, making them suitable for applications where functional data is involved with all components of the causal mediation framework. Currently, the scope of this area of study is limited to linear modelling of the mediation framework. Zero-inflated count data, observed in many applications across fields such as healthcare, ecology, and economics, involves count outcomes with an excess of zeros. For instance, the number of hospital readmissions, counts of dental caries where many individuals may have no cavities, the number of substance use relapses where some individuals remain abstinent, and insurance claims where no claims are made during the study period all exhibit this pattern. Causal mediation analysis with zero-inflated count outcomes entails a nonlinear relationship between the treatment, mediator and outcome. Related works have been done in a scalar setting \citep{wangzero, chengzero}. Non-functional causal mediation analysis with a zero-inflated count outcome typically involves fitting a linear model for the mediator and a zero-inflated count model, such as the zero-inflated Poisson or zero-inflated Negative Binomial models, for the outcome. However, such problems have yet to be examined under scenarios of a functional mediator. Another important limitation of the existing literature is its predominant focus on densely observed functional data, which is impractical in many practical scenarios. In contrast, infrequently and irregularly observed functional mediators are prevalent in applications such as electronic health records and behavioural studies, where continuous records are infeasible or adherence to study protocols varies across individuals. When observations of the individual trajectories happen at sparsely distributed time points and vary among subjects, standard FDA methods can result in poorly recovered trajectories. These gaps motivate the development of causal mediation methods that accommodate both zero-inflation and sparse observation schemes.

Research on nonlinear and sparse FDA techniques has emerged in the recent decade \citep{revfda}. In the context of zero-inflated scalar outcomes, \citet{fzip} attempts to model hospitalisations in patients with dialysis by proposing functional linear models for zero-inflated count data. Their work outlined the formulation and estimation of a functional zero-inflated Poisson (ZIP) model with a functional predictor and multiple cross-sectional predictors to model counts generated by a mixture distribution. To address infrequent and irregular observation designs, functional principal component analysis (FPCA) has been extended through principal component analysis through conditional expectation (PACE) by \citet{yao2005sparse}, which estimates mean and covariance functions via pooled local smoothing and recovers individual trajectories through a linear mixed model. This approach not only reconstructs sparsely observed curves but also reduces dimensionality via functional principal component (FPC) scores, facilitating downstream analysis. Together, these developments provide a foundation for modelling strategies that accommodate both zero-inflated outcomes and sparse functional mediators.

Motivated by the research question, we extend existing causal mediation methodologies by introducing a framework that incorporates both a sparse functional mediator and a zero-inflated count outcome. The potential outcomes framework \citep{rubin1974random} is employed to define the causal effects of interest in this context and provide the theoretical underpinning for our approach, including conditions for effect identification. To address both the sparse functional mediator and the zero-inflated count outcomes, functional linear and nonlinear models are implemented. Estimation and inference on the direct and indirect effects are performed by a parameter-simulating quasi-Bayesian Monte Carlo approximation method based on the mediation formula \citep{pearl2012causal}. Simulation studies validate our approach, demonstrating its capability to estimate causal effects reliably in this context. Using the method proposed, we find both a significant total effect of gender on rehospitalisation counts and a significant natural indirect effect channelled through time-varying CVP.

In this paper, we start by introducing the causal mediation analysis with a sparse functional mediator and a zero-inflated count outcome using the potential outcomes framework in Section \ref{sec2}. We define the causal total effect, natural direct effect, and natural indirect effect, as well as present assumptions and identification of the causal effects. In Section \ref{sec3}, we demonstrate the proposed methods, including the mediator model, the outcome model, a parameter-simulating quasi-Bayesian Monte Carlo approximation algorithm, and a sensitivity analysis for potential unmeasured mediator-outcome confounding. The performances of the proposed method are assessed via simulation studies in Section \ref{sec4}. Section \ref{sec5} addresses the role of gender as the treatment in a causal question, describes the data selected from MIMIC-IV, and discusses findings regarding our research question. Section \ref{sec6} concludes the paper by giving a discussion.

\section{Sparse functional causal mediation analysis with a zero-inflated outcome}\label{sec2}

\tikzset{
    box/.style={rectangle, draw, text centered, minimum height=2em},
    path/.style={->,>=stealth,thick}
}
\begin{figure}[!t]
\centering
\begin{tikzpicture}[node distance=4.5cm]
\node[box] (exposureB) {Treatment $A$};
\node[box, right of=exposureB] (mediatorB) {Functional Mediator $M(t)$};
\node[box, right of=mediatorB] (outcomeB) {Outcome $Y$};
\node[box] (covB) at ($(exposureB)!0.5!(outcomeB) + (0,2cm)$) {Covariates $\mathbf{X}$};
\draw[path] (exposureB) -- node[sloped, above] {} (mediatorB);
\draw[path] (mediatorB) -- node[sloped, above] {} (outcomeB);
\draw[path] (exposureB) |- ([yshift=-6mm]exposureB.south) -| (outcomeB);
\draw[path] (covB) -- node[below] {} (mediatorB);
\draw[path] (covB) -- node[below] {} (exposureB);
\draw[path] (covB) -- node[below] {} (outcomeB);
\end{tikzpicture}
\caption{A causal diagram for a causal mediation analysis with a functional mediator.}
\label{fig:causaldiagram}
\end{figure}

As illustrated in Figure \ref{fig:causaldiagram}, the causal mediation analysis of interest involves a binary treatment $A$, a zero-inflated count outcome $Y$, a $p_x$-dimensional vector of baseline covariates $\mathbf{X}$, and a time-varying mediator $M(t)$ as a function of time. Without loss of generality, we rescale the time $t$ so that $t \in [0,1]$. We explore a study with a sample of $n$ individuals, where each pertains to either the treatment group with $A_i=1$, or the control group with $A_i=0$, for $i=1,\ldots,n$. The functional mediator $M(t)$ is a function of time and is therefore a smooth stochastic process. For each individual $i$, discrete observations $\Tilde{M}_i(t_{ij})$ of the mediator occur at $T_i$ irregularly spaced time points $t_{ij}$ with $j=1,\ldots,T_i$, and are regarded as realisations, potentially contaminated with measurement errors, of the underlying smooth process. We assume that the end-point outcome $Y_i$ is measured at the end of the study. The baseline covariates $\mathbf{X}$ include variables that confound the relation between $A$ and $Y$, the relation between $M(t)$ and $Y$, or the relation between $A$ and $M(t)$, and are not affected by $A$. Both the treatment assignment and baseline covariate measurement take place at the beginning of the study.

The causal estimands are defined in Section \ref{subsec2.1} followed by a discussion on the relevant identifying assumptions in Section \ref{subsec2.2}. In \ref{subsec2.3}, we present nonparametric identification results of the causal effects.

\subsection{Definition of causal effects}\label{subsec2.1}

In defining the causal effects of interest, we introduce the potential outcomes framework in the context of mediation analysis with a functional mediator. To accommodate the functional mediator, the bold font notation $\mathbf{M} = \{M(t) \mid t \in [0,1]\}$ is used to denote the mediator process over the entire time range. It is worth pointing out that $\mathbf{M}$ has realisations as sample paths $\mathbf{m} \in \mathcal{D}_\mathbf{M}^{[0,1]}$, where $\mathcal{D}_\mathbf{M}^{[0,1]} \subseteq \mathbb{R}^{[0,1]}$ refers to the range of the mediator process and contains real-valued functions defined on the interval $[0,1]$.  We can consequently write $\mathbf{M}_i(a)$, for $a = 0, 1$, as the potential values of the underlying mediator process over the entire time range for individual $i$ under treatment level $a$. $Y_i(a, \mathbf{M}_i(a^\prime))$ then refers to the potential value of the outcome $Y$ for individual $i$ if, potentially counterfactually, the individual received treatment $a$ and had values of the mediator process at the level that would have been observed under treatment $a^\prime$.

We can then define the total effect (TE) of the treatment $A$ on the outcome $Y$ as
\begin{equation}
\tau_{\textsubscript{TE}} \equiv E\left\{Y\left(1, \mathbf{M}(1)\right) - Y\left(0, \mathbf{M}(0)\right)\right\}.\label{eq:totaleffect}
\end{equation}
The total effect is defined as the mean difference in potential outcomes under treatment levels 1 and 0. \citet{imaimed} proposes a framework for decomposing the total effect (TE) into direct and indirect effects in the presence of a mediator. We build upon this framework by extending it to accommodate a functional mediator, allowing for definitions of the natural direct and indirect effects in our settings. In particular, the natural indirect effect (NIE) is defined as the mean difference in the potential outcomes under a fixed treatment level but varied potential values of the mediator process under different treatment levels,
\begin{equation}
\tau_{\textsubscript{NIE}}(a) \equiv E\left\{Y\left(a, \mathbf{M}(1)\right) - Y\left(a, \mathbf{M}(0)\right)\right\}, \quad a=0,1.\label{eq:nie}
\end{equation}
The NIE measures the component of the treatment's causal effect on the outcome channelled through the mediator. Similarly, the natural direct effect (NDE) is the effect that does not go through the mediator and is defined as the mean difference in the potential outcomes under different treatment values and the potential values of the mediator process at a fixed treatment level,
\begin{equation}
\tau_{\textsubscript{NDE}}(a) \equiv E\left\{Y\left(1, \mathbf{M}(a)\right) - Y\left(0, \mathbf{M}(a)\right)\right\}, \quad a=0,1.\label{eq:nde}
\end{equation}
The TE can be decomposed as the sum of NIE and NDE,
\begin{equation*}
\tau_{\textsubscript{TE}} = \tau_{\textsubscript{NIE}}(a) + \tau_{\textsubscript{NDE}}(1-a), \quad a=0,1. 
\end{equation*}

The definitions of the causal estimands of interest require comparing four potential outcomes: $Y\left(1, \mathbf{M}(1)\right)$, $Y\left(1, \mathbf{M}(0)\right)$, $Y\left(0, \mathbf{M}(1)\right)$, and $Y\left(0, \mathbf{M}(0)\right)$. However, in practice, only one of these potential outcomes can be observed, which is also known as the fundamental problem of causal inference \citep{holland1986statistics}. As a result, certain assumptions are needed in order to identify the potential outcomes from the observed data.

\subsection{Identifying assumptions}\label{subsec2.2}

The assumptions of identification for causal mediation analysis have been intensively discussed \citep{petersen2006estimation, medform, vanderweele2014effect, pearl2022direct, Nguyenmedassump}. However, such assumptions involving a functional mediator need further clarification. Following the guidance of \citet{Nguyenmedassump} and \citet{zhaou2024unpublished}, we examine and outline the identifying assumptions for causal mediation analysis with a functional mediator. 

The identification assumptions listed below are analogous to those commonly used in causal mediation analysis with scalar mediators. However, when the mediator takes the form of a continuous-time stochastic process, these assumptions need adaptations to account for the functional nature of the data. Specifically, the mediator now lies in an infinite-dimensional function space, and the assumptions must be formulated over Borel measurable subsets of the range of the mediator process $\mathcal{D}_\mathbf{M}^{[0,1]}$. Moreover, assumptions involving consistency and cross-world potential outcomes should be interpreted over entire sample paths of the mediator process, rather than pointwise evaluations.

\begin{assumption}[Positivity]
For $a \in \left\{0, 1\right\}$, $P(A = a \mid \mathbf{X}) > 0$; for any Borel measurable set $\mathcal{B} \subseteq \mathcal{D}_\mathbf{M}^{[0,1]}$ that has positive measures, the probability that a sample path $\mathbf{m}$ of the process $\mathbf{M}$ belongs to $\mathcal{B}$ given $\mathbf{X}$ and $A = a$ is positive: $P(\mathbf{m} \in \mathcal{B} \mid \mathbf{X}, A = a) > 0$. 
\end{assumption}

\begin{assumption}[Consistency of the potential outcome]
Given any Borel measurable set $\mathcal{B} \subseteq \mathcal{D}_\mathbf{M}^{[0,1]}$ that has positive measures, for all $\mathbf{m} \in \mathcal{B}$ and $a \in \left\{0, 1\right\}$, if $A = a$ and $\mathbf{M} = \mathbf{m}$,
\begin{equation*}
    Y = Y(a, \mathbf{m}).
\end{equation*}
Namely, the observed outcome reveals the potential outcome under the respective treatment level and mediator process.
\end{assumption}

\begin{assumption}[Consistency of the potential mediator]
For $a \in \left\{0, 1\right\}$, if $A = a$,
\begin{equation*}
    \mathbf{M} = \mathbf{M}(a).
\end{equation*}
Namely, the observed mediator process reveals the potential mediator process under the respective treatment level.
\end{assumption}

\begin{assumption}[Consistency of the cross-world potential outcome]
Given any Borel measurable set $\mathcal{B} \subseteq \mathcal{D}_\mathbf{M}^{[0,1]}$ that has positive measures, for all $\mathbf{m} \in \mathcal{B}$ and $a, a^\prime \in \left\{0, 1\right\}$, if $\mathbf{M}(a^\prime) = \mathbf{m}$,
\begin{equation*}
    Y(a, \mathbf{m}) = Y(a, \mathbf{M}(a^\prime)).
\end{equation*}
Namely, the potential outcome reveals the cross-world potential outcome under the respective potential mediator process.
\end{assumption}

\begin{assumption}[Conditional independence] \label{assump:conditionalindep}
Given any Borel measurable set $\mathcal{B} \subseteq \mathcal{D}_\mathbf{M}^{[0,1]}$ that has positive measures, for all $\mathbf{m} \in \mathcal{B}$ and $a, a^\prime \in \left\{0, 1\right\}$,
\begin{align*}
    \left\{Y(a^\prime , \mathbf{m}), \mathbf{M}(a)\right\} \ind A &\mid \mathbf{X},\\
    Y(a^\prime , \mathbf{m}) \ind \mathbf{M}(a) &\mid \mathbf{X}, A = a.
\end{align*}
Namely, given the observed baseline covariates, the treatment is unconfounded for the potential outcomes and potential mediator processes, and the mediator process is unconfounded for the potential outcomes.
\end{assumption}

\subsection{Nonparametric identification of the causal effects}\label{subsec2.3}
Under the previously discussed identifying assumptions, the causal effects in a causal mediation analysis involving a functional mediator can be nonparametrically identified from the observed data distribution. Specifically,
\begin{align*}
    \tau_{\textsubscript{NIE}}(a) &= E_{\mathbf{X}}\left\{E\left(Y \mid A=a, \mathbf{X}\right) \right\} -  E_{\mathbf{X}}\left[E_{\mathbf{M} \mid A = a^\prime, \mathbf{X}} \left\{E\left(Y \mid A = a, \mathbf{M}, \mathbf{X}\right)\right\}\right],\\
    \tau_{\textsubscript{NDE}}(a^\prime) &= E_{\mathbf{X}}\left[E_{\mathbf{M} \mid A = a^\prime, \mathbf{X}} \left\{E\left(Y \mid A = a, \mathbf{M}, \mathbf{X}\right)\right\}\right] - E_{\mathbf{X}}\left\{E\left(Y \mid A=a^\prime, \mathbf{X}\right) \right\}, \\
    \tau_{\textsubscript{TE}} &= E_{\mathbf{X}}\left\{E\left(Y \mid A=a, \mathbf{X}\right) \right\} -  E_{\mathbf{X}}\left\{E\left(Y \mid A=a^\prime, \mathbf{X}\right) \right\}.
\end{align*}
Consequently, to estimate the causal effects, the nonparametric identification results require modelling: (a) the observed outcome $Y$ conditional on the observed treatment $A$, mediator process $\mathbf{M}$ and baseline covariates $\mathbf{X}$; (b) the observed mediator process $\mathbf{M}$ conditional on the observed treatment $A$ and baseline covariates $\mathbf{X}$. 

\section{Methods proposed}\label{sec3}

Without the presence of functional data and nonlinearity, the models mentioned in Section \ref{subsec2.3} correspond to the linear structural equation models (SEM) used in traditional mediation analysis \citep{baron1986moderator}. Causal mediation analysis, by defining the causal effects using the potential outcomes frameworks, enables the use of more complex statistical models to accommodate a functional mediator and a zero-inflated count outcome. We describe a function-on-scalar regression for the sparse functional mediator and its estimation via a mixed model representation in Section \ref{subsec3.1}. Section \ref{subsec3.2} presents a functional zero-inflated Poisson (ZIP) model for the outcome. Finally, we propose a parameter-simulating quasi-Bayesian Monte Carlo approximation method to perform estimation and inference for the causal effects in Section \ref{subsec3.3}.

\subsection{A function-on-scalar regression mediator model}\label{subsec3.1}

\subsubsection{Model formulation}
We propose to model the smooth mediator process underlying the infrequent and irregular mediator measurements with a function-on-scalar regression \citep{ramsay2005fdabook}. Specifically, it is assumed that the functional mediator $M(t)$ follows a function-on-scalar regression on both the treatment $A$ and the baseline covariates $\mathbf{X}$,
\begin{equation}\label{eq:mediatormodel}
    M_{i}(t) = \beta_{0}(t) + \beta_{1}(t)A_i +  \sum_{d=1}^{p_x} \beta_{2d}(t)X_{id} + \varepsilon_{i}(t), \quad \text{for} \quad t \in [0,1].
\end{equation} 
The intercept function is $\beta_0(t)$, the functional coefficients $\beta_{1}(t)$ and $\beta_{2d}(t)$ represent the partial effects of treatment $A$ and covariate $X_{d}$ on the mediator process $M(t)$ at time $t$, and $\varepsilon_{i}(t)$ is a Gaussian random error process with mean zero.

\subsubsection{Estimation via mixed model representation}

One important caveat with estimating Model (\ref{eq:mediatormodel}) is the fact that the mediator process is infrequently and irregularly observed with potential measurement errors. Practically, everything related to the underlying mediator process $M_i(t)$ is unknown. Instead, discrete observations $\Tilde{M}_i(t_{ij})$ are assumed to happen at $T_i$ observation times, denoted as $t_{i1}, \ldots, t_{iT_i}$, for the $i$-th individual. We also assume the underlying mediator process is observed with measurement errors $\epsilon_{ij}$ where $\epsilon_{ij} \stackrel{\text{iid}}{\sim} N(0, \sigma^2_M)$. This means the discrete observations $\Tilde{M}_i(t_{ij})$ from the data can be represented as
\begin{equation}\label{eq:observedmediatorwitherror}
    \Tilde{M}_i(t_{ij}) = M_i(t_{ij}) + \epsilon_{ij}.
\end{equation}

\citet{scheipl2015functional} presents a flexible functional additive mixed model framework that accommodates a potentially sparsely measured functional response and scalar or functional covariates. Coefficient functions are represented using a set of pre-determined cubic spline bases, which facilitates estimating model parameters via penalised maximum likelihood under a linear mixed model framework.

Specifically, each coefficient function is expanded using $K_b$ cubic spline basis functions $\left\{\phi_1(t), \cdots, \phi_{K_b}(t)\right\}$, where a relatively large $K_b$ is chosen to provide sufficient modelling flexibility. Specifically,
\begin{align*}
    \beta_0(t)&=\sum_{k_b=1}^{K_b} b_{0 k_b} \phi_{k_b}(t) = \boldsymbol{\phi}(t)^\top\mathbf{b}_0, \\
    \beta_1(t)&=\sum_{k_b=1}^{K_b} b_{1 k_b} \phi_{k_b}(t) = \boldsymbol{\phi}(t)^\top\mathbf{b}_1, \\
    \beta_{2d}(t) &= \sum_{k_b=1}^{K_b} \mathbf{b}_{2dk_b} \phi_{k_b}(t) = \boldsymbol{\phi}(t)^\top\mathbf{b}_{2d},
\end{align*}
for $d = 1, \ldots, p_x$, where $\boldsymbol{\phi}(t)$ is a vector of functions containing the $K_b$ cubic spline basis functions and $\mathbf{b}_0, \mathbf{b}_1, \mathbf{b}_{2d}$ are vectors of $K_b$ basis coefficients corresponding to each of the coefficient functions. Model (\ref{eq:mediatormodel}) can thus be written as
\begin{equation} \label{eq:reducedmediatormodel}
    M_i(t) = \boldsymbol{\phi}(t)^\top\mathbf{b}_0 + A_i\boldsymbol{\phi}(t)^\top\mathbf{b}_1 + \sum_{d=1}^{p_x} X_{id} \boldsymbol{\phi}(t)^\top\mathbf{b}_{2d} + \varepsilon_i(t).
\end{equation}
Combined with (\ref{eq:observedmediatorwitherror}), the raw observations of the mediator process admit the form
\begin{equation} \label{eq:observedmediatormodel}
    \Tilde{M}_i(t_{ij}) = \boldsymbol{\phi}(t_{ij})^\top\mathbf{b}_0 + A_i\boldsymbol{\phi}(t_{ij})^\top\mathbf{b}_1 + \sum_{d=1}^{p_x} X_{id} \boldsymbol{\phi}(t_{ij})^\top\mathbf{b}_{2d} + E_{ij},
\end{equation}
where $E_{ij} = \varepsilon_i(t_{ij}) + \epsilon_{ij}$. Now, let $\mathcal{M} \in \mathbb{R}^{N}$ denote a vector of length $N = \sum^{n}_{i = 1} T_i$, containing all observed mediator values $\Tilde{M}_i(t_{ij})$ stacked across individuals and time points. We construct a design matrix $\boldsymbol{\Phi} \in \mathbb{R}^{N \times O}$, with $O = K_b(p_x+2)$, where each row corresponds to one observation $\Tilde{M}_i(t_{ij})$ and takes the form
\[
\left[\boldsymbol{\phi}(t_{ij})^\top, A_i \boldsymbol{\phi}(t_{ij})^\top, X_{i1} \boldsymbol{\phi}(t_{ij})^\top, \dots, X_{ip_x} \boldsymbol{\phi}(t_{ij})^\top \right]
\]
Let \( \mathbf{b} = [\mathbf{b}_0^\top, \mathbf{b}_1^\top, \mathbf{b}_{21}^\top, \dots,\ \mathbf{b}_{2p_x}^\top]^\top \) denote the stacked vector of all basis coefficients and $\boldsymbol{E}$ denote the stacked vector containing the errors $E_{ij}$, Model (\ref{eq:observedmediatormodel}) can be written in matrix form as
\[
\mathcal{M} = \boldsymbol{\Phi} \mathbf{b} + \boldsymbol{E}.
\]
To prevent overfitting due to the high flexibility of the selected spline basis system, a roughness penalty is imposed on the coefficient functions. This leads to a penalised likelihood estimation problem, where the objective function to be minimised is
\begin{equation}\label{eq:penalisedlikelihood}
\mathcal{L}_{\mathrm{pen}}(\mathbf{b})=\frac{1}{\sigma_E^2}\|\mathcal{M}-\boldsymbol{\Phi} \mathbf{b}\|^2+\frac{\delta_0}{\sigma_E^2} \mathbf{b}_0^{\top} \mathbf{P} \mathbf{b}_0+\frac{\delta_1}{\sigma_E^2} \mathbf{b}_1^{\top} \mathbf{P} \mathbf{b}_1+\sum_{d=1}^{p_x} \frac{\delta_{2 d}}{\sigma_E^2} \mathbf{b}_{2 d}^{\top} \mathbf{P} \mathbf{b}_{2 d},
\end{equation}
where $\sigma_E^2$ is the variance of the errors $E_{ij}$, $\delta_0$, $\delta_1$ and  $\delta_{2d}$ are smoothing parameters that control the balance between goodness of fit and roughness of the estimated coefficient functions, and $\mathbf{P} = \int_0^1 \boldsymbol{\phi}^{(2)}(t)\boldsymbol{\phi}^{(2)}(t)^\top $ is a $K_b \times K_b$ the penalty matrix constructed to penalise the integrated squared second derivative of each coefficient function. This penalised spline formulation is equivalent to a linear mixed effects model in which the spline basis coefficients $\mathbf{b}_0, \mathbf{b}_1, \mathbf{b}_{2d}$ are treated as random effects \citep{ruppert2003semiparametric}. Each penalty term in \eqref{eq:penalisedlikelihood}, such as $\frac{\delta_0}{\sigma_E^2} \mathbf{b}_0^\top \mathbf{P} \mathbf{b}_0$, is equivalent to assuming a Gaussian prior $\mathbf{b}_0 \sim N(\mathbf{0}, \zeta_0 \mathbf{P}^{-1})$, where $\zeta_0 = \sigma_E^2 / \delta_0$. Similarly, $\mathbf{b}_1$ and each of $\mathbf{b}_{2d}$ are treated as random effects with variances $\zeta_1 = \sigma_E^2 / \delta_1$ and $\zeta_{2d} = \sigma_E^2 / \delta_{2d}$, respectively.  Then, the solution $\widehat{\mathbf{b}}$ to (\ref{eq:penalisedlikelihood}) is the best linear unbiased predictor (BLUP). The smoothing parameters are consequently estimated through restricted maximum likelihood (REML) under the additive model framework using the \texttt{pffr} function in the \texttt{R} package \texttt{refund} \citep{refundpack}. According to \citet{reiss2009smoothing} and \citet{wood2011fast}, REML typically produces more stable estimates than methods based on generalised cross-validation (GCV). It is also more robust to misspecification of the error correlation structure and performs reliably in a broad range of practical settings \citep{krivobokova2007note}.

\subsection{A functional zero-inflated Poisson outcome model}\label{subsec3.2}

\subsubsection{Model formulation}

While most existing literature on rehospitalisation focuses on the rate of hospital readmissions, our study examines the number of readmissions within 60 days following discharge from the primary CABG hospital stay. This outcome provides deeper insights, albeit posing greater modelling challenges due to zero-inflation. The zero-inflated Poisson (ZIP) model, proposed by \citet{lambertzip}, models count data with excessive zero-valued observations by considering a mixture of a binary model for generating zeros and a standard Poisson model for generating non-zero counts, as well as some zeros. Zeros produced by the former are often referred to as ``structural zeros'' or ``excess zeros'' since the process cannot generate any counts other than zero. \citet{fzip} later extended it to a functional version, with which we can parameterise the ZIP model with functional covariates. We assume that the end-point outcome $Y$ follows a functional ZIP model on the underlying mediator process, treatment and baseline covariates,
\begin{equation} \label{eq:outcomemodel}
    P(Y_i = y_i | \mathbf{M}_i, A_i, \mathbf{X}_i) = 
    \begin{cases}
    p_i + (1 - p_i)e^{-\theta_i}, & \text{if } y_i = 0 \\
    (1 - p_i) \frac{e^{-\theta_i} \theta_i^{y_i}}{y_i!}, & \text{if } y_i > 0
    \end{cases},
\end{equation}
where $p_i$ and $\theta_i$ are related to the functional mediator, the treatment, and the baseline covariates through some link functions. Here we consider a logit link function for $p_i$ and a log link function for $\theta_i$ where
\begin{align}
    \log\frac{p_i}{1-p_i} &= \alpha_0 + \int_0^1 \alpha_1(t)M_i(t) \, dt + \alpha_2A_i + \mathbf{X}_i^\top\boldsymbol{\alpha}_3, \label{eq:logisticlink}\\
    \log\theta_i &= \gamma_0 + \int_0^1 \gamma_1(t)M_i(t) \, dt + \gamma_2A_i + \mathbf{X}_i^\top \boldsymbol{\gamma}_3. \label{eq:loglink}
\end{align}
The individual end-point outcome $Y_i$ is either an excess zero, which takes the value of zero from a Bernoulli distribution with probability $p_i$, or a zero from a Poisson distribution that has mean $\theta_i$ with probability $1-p_i$.

\subsubsection{Estimation via principal component analysis through conditional expectation}
Estimation of Model (\ref{eq:outcomemodel}) requires jointly estimating the parameters in the log-linear model for $\theta_i$ and the logistic model for $p_i$. The problem is thus to estimate the scalar-on-function models for $\log(\frac{p_i}{1-p_i})$ and $\log\theta_i$. To transform the scalar-on-function models into their scalar analogues, we use PACE to recover the mean function, eigenfunctions and scores of the underlying mediator process $M_i(t)$ \citep{yao2005sparse}. This procedure serves two primary purposes: (a) recovering the underlying mediator process by pooled local linear smoothing and best linear unbiased prediction; and (b) decomposing the estimated mediator process into a finite number of FPCs for subsequent maximum likelihood estimation of the finite-dimensional ZIP model.

The proposed functional ZIP model concerns the unobservable mediator process $M_i(t), t \in [0, 1]$ underlying the discrete infrequently and irregularly observed mediator $\Tilde{M}_i(t_{ij})$ for individual $i$ at observation times $t_{ij}$, where $i = 1,\ldots,n$ and $j = 1,\ldots,T_i$. For each underlying process $M_i(t)$, we define a bivariate function $C\left(s, t\right)=\operatorname{Cov}\left[M_i(s), M_i\left(t\right)\right]$ that quantifies the covariance between its values at two time points, known as its covariance operator. The positive definite function $C\left(s, t\right)$ admits a sequence of positive eigenvalues denoted by $\lambda_1>\lambda_2>\cdots>0$, and a set of orthonormal eigenfunctions denoted by $\psi_1, \psi_2, \ldots$, such that $C\left(s, t\right) = \sum_{k = 1}^\infty \lambda_k\psi_k(s)\psi_k(t)$. With the above spectral decomposition, the Karhunen–Lo\'{e}ve expansion states that $M_i(t)$ can be represented as
\begin{equation*}
M_i(t)=\mu(t) + \sum_{k=1}^{\infty} \xi_{ik} \psi_k(t),
\end{equation*}
where $\mu(t) = E\left[M(t)\right]$ is the mean function, $\xi_{ik}=\int_0^1\left\{M_i(t)-\mu(t)\right\} \psi_k(t) d t$, called the functional principal component (FPC) scores, are uncorrelated random variables with mean $0$ and variance $\lambda_k$. The FPC scores $\xi_{ik}$ quantify the extent to which individual trajectories deviate from the mean function $\mu(t)$ along the direction of the corresponding eigenfunctions $\psi_k(t)$, providing an interpretable summary of the underlying functional variability. Discrete observations $\Tilde{M}_i(t_{ij})$ are assumed to be measured potentially with errors as in (\ref{eq:observedmediatorwitherror}) and so can be written as
\begin{equation}\label{eq:observedmediatordecomp}
\Tilde{M}_i(t_{ij}) = M_i(t_{ij}) + \epsilon_{ij} = \mu(t_{ij}) + \sum_{k=1}^{\infty} \xi_{ik} \psi_k(_{ij}) + \epsilon_{ij}, \quad t_{ij} \in [0,1],\ i = 1,\ldots, n,\ j = 1,\ldots, T_i.
\end{equation}

PACE estimates the mean function $\mu(t)$ and covariance function $C(s, t)$ of the underlying mediator process $M_i(t)$ by pooled local linear smoothing. The estimated eigenvalues $\widehat{\lambda}_k$ and orthonormal eigenfunctions $\widehat{\psi}_k(t)$ are obtained by eigenanalysis of the estimated covariance surface. Another decision is to choose a finite number of FPCs, $K$, to truncate (\ref{eq:observedmediatordecomp}). Again, different approaches exist for an estimate $\widehat{K}$ from the data \citep{yao2005sparse}, we choose $\widehat{K}$ based on the fraction of variance explained and set the threshold to $90\%$. The measurement error variance $\sigma_M^2$ is estimated by taking a fraction of the difference between the diagonal elements of the discretised estimated covariance matrix and a transformed version of it by rotating the coordinates in local linear smoothing. Technical details are omitted as they closely follow \citet{yao2005sparse}.

The key novelty of \citet{yao2005sparse} is to estimate the FPC scores from a linear mixed model based on the $K$-truncated Karhunen–Lo\'{e}ve expansion
\begin{equation} \label{mixedmodelscores}
\Tilde{M}_i(t_{ij})=\mu\left(t_{i j}\right)+\sum_{k=1}^K \xi_{i k} \psi_k\left(t_{i j}\right)+\epsilon_{ij}, \quad \boldsymbol{\xi}_i \stackrel{\text{iid}}{\sim} N(\mathbf{0}, \boldsymbol{\Lambda}),\ \epsilon_{ij} \stackrel{\text{iid}}{\sim} N(0, \sigma^2_M), 
\end{equation}
where $\boldsymbol{\xi}_i = \left(\xi_{i1}, \ldots, \xi_{ik}\right)^\top$ is a vector of length $K$ and is assumed to be independent of the measurement errors. Notationally, we represent the collection of FPC decomposition objects with $\boldsymbol{\varphi} = \left\{\boldsymbol{\mu}, \boldsymbol{\psi}, \boldsymbol{\Lambda}, K, \sigma_M^2\right\}$, where $\boldsymbol{\mu}$ is the mean function evaluated over $[0,1]$, $\boldsymbol{\psi}$ are the $K$ eigenfunctions evaluated over $[0,1]$, and $\boldsymbol{\Lambda} = \operatorname{diag}\left\{\lambda_1, \ldots \lambda_K\right\}$ is a diagonal matrix of eigenvalues. Let $\boldsymbol{t}_i = \left\{t_{ij}\right\}_{j = 1}^{T_i}$ be the observation grid for individual $i$, $\Tilde{\boldsymbol{M}}_i(\boldsymbol{t}_i) = \left(\Tilde{M}_i(t_{i1}), \ldots, \Tilde{M}_i(t_{ij})\right)^\top$ be a vector of discrete observations of the mediator for individual $i$, plugging in the estimates $\widehat{\boldsymbol{\varphi}} = \left\{\widehat{\boldsymbol{\mu}}, \widehat{\boldsymbol{\psi}}, \widehat{\boldsymbol{\Lambda}}, \widehat{K}, \widehat{\sigma}_M^2\right\}$ into Model (\ref{mixedmodelscores}) yields the estimated BLUPs for the FPC scores
\begin{equation}\label{eq:estimatedscores}
\widehat{\boldsymbol{\xi}}_i= \mathrm{E}\left[\boldsymbol{\xi}_i \mid \Tilde{\boldsymbol{M}}_i(\boldsymbol{t}_i), \widehat{\boldsymbol{\varphi}}\right]=\left(\widehat{\boldsymbol{\psi}}\left(\boldsymbol{t}_i\right)^\top \widehat{\boldsymbol{\psi}}\left(\boldsymbol{t}_i\right)+\widehat{\sigma}^2_M \widehat{\boldsymbol{\Lambda}}^{-1}\right)^{-1} \widehat{\boldsymbol{\psi}}\left(\boldsymbol{t}_i\right)^\top\left(\Tilde{\boldsymbol{M}}_i(\boldsymbol{t}_i)-\widehat{\boldsymbol{\mu}}\left(\boldsymbol{t}_i\right)\right),
\end{equation}
where $\widehat{\boldsymbol{\psi}}\left(\boldsymbol{t}_i\right)$ is a $T_i \times \widehat{K}$ matrix consisting of the estimated eigenfunctions evaluated over the grid $\boldsymbol{t}_i$, and $\widehat{\boldsymbol{\mu}}\left(\boldsymbol{t}_i\right)$ is a vector of length $T_i$ containing the estimated mean function evaluated over the grid $\boldsymbol{t}_i$. Now consider $\boldsymbol{t}_g$ as the union of $\boldsymbol{t}_i$ of all individuals, which is usually a dense grid on the interval $[0,1]$. Using the estimated FPC scores from (\ref{eq:estimatedscores}), the estimated underlying mediator process evaluated on $\boldsymbol{t}_g$ is given by
\begin{equation}\label{eq:estimatedcurve}
\widehat{\boldsymbol{M}}_i\left(\boldsymbol{t}_g\right)=\mathrm{E}\left[\Tilde{\boldsymbol{M}}_i(\boldsymbol{t}_g) \mid \widehat{\boldsymbol{\xi}}_i, \widehat{\boldsymbol{\varphi}}\right] =\widehat{\boldsymbol{\mu}}\left(\boldsymbol{t}_g\right)+\sum_{k=1}^{\widehat{K}} \widehat{\xi}_{i k} \widehat{\psi}_k\left(\boldsymbol{t}_g\right) =\widehat{\boldsymbol{\mu}}\left(\boldsymbol{t}_g\right)+\widehat{\boldsymbol{\psi}}\left(\boldsymbol{t}_g\right) \widehat{\boldsymbol{\xi}}_{i}.
\end{equation}
We carry out the PACE procedures using the \texttt{FPCA} function in the \texttt{R} package \texttt{fdapace}. The end product, Equation (\ref{eq:estimatedcurve}), achieves the two goals of an FPCA for the sparse longitudinal mediator: (a) $\widehat{\boldsymbol{M}}_i\left(\boldsymbol{t}_g\right)$ on a dense grid $\boldsymbol{t}_g$, as an estimate of the smooth mediator process underlying the error-contaminated sparse observations; (b) $\widehat{\boldsymbol{\mu}}\left(\boldsymbol{t}_g\right)$ and $\widehat{\boldsymbol{\psi}}\left(\boldsymbol{t}_g\right) \widehat{\boldsymbol{\xi}}_{i}$ for a finite $\widehat{K}$-truncated Karhunen–Lo\'{e}ve decomposition.

Consequently, with the estimated mean function, eigenfunctions and scores from PACE, Models (\ref{eq:logisticlink}) and (\ref{eq:loglink}) can be written as
\begin{align}
    \log\frac{p_i}{1-p_i} & \approx \alpha^*_0+\sum_{k=1}^{\widehat{K}} \widehat{\xi}_{i k} \alpha^*_k+\alpha_2A_i + \mathbf{X}^\top_i\boldsymbol{\alpha}_3 \label{eq:reducedlogisticlink}\\
    \log\theta_i & \approx \gamma^*_0+\sum_{k=1}^{\widehat{K}} \widehat{\xi}_{i k} \gamma^*_k+\gamma_2A_i + \mathbf{X}^\top_i\boldsymbol{\gamma}_3, \label{eq:reducedloglink}
\end{align}
where
\begin{align*}
&\alpha^*_0=\alpha_0+\int_0^1 \alpha_1(t) \widehat{\mu}(t) d t, \quad \alpha^*_k=\int_0^1 \alpha_1(t) \widehat{\psi}_k(t) d t \quad \text { and } \\
&\gamma^*_0=\gamma_0+\int_0^1 \gamma_1(t) \widehat{\mu}(t) d t, \quad \gamma^*_k=\int_0^1 \gamma_1(t) \widehat{\psi}_k(t) d t,    
\end{align*}
The parameters $\boldsymbol{\alpha} = (\alpha^*_0, \alpha^*_1, \ldots, \alpha^*_{\widehat{K}}, \alpha_2, \boldsymbol{\alpha}^\top_3)^\top$ and $\boldsymbol{\gamma} = (\gamma^*_0, \gamma^*_1, \ldots, \gamma^*_{\widehat{K}}, \gamma_2, \boldsymbol{\gamma}^\top_3)^\top$ from Models (\ref{eq:reducedlogisticlink}) and (\ref{eq:reducedloglink}) can then be estimated by maximising the log-likelihood
\begin{align}
        \ell\left(\boldsymbol{\alpha}, \boldsymbol{\gamma}\right)=& \sum_{Y_i=0} \log \left[p_i\left(\boldsymbol{\alpha}\right) +\left\{1-p_i\left(\boldsymbol{\alpha}\right)\right\}e^{-\theta_i\left(\boldsymbol{\gamma}\right)}\right] \nonumber \\
& +\sum_{Y_i>0}\left[\log \left\{1 - p_i\left(\boldsymbol{\alpha}\right)\right\}-\theta_i\left(\boldsymbol{\gamma}\right)
+Y_i \log \left\{\theta_i\left(\boldsymbol{\gamma}\right)\right\}-\log \left(Y_{i} !\right)\right].
\end{align}

\subsection{A parameter-simulating quasi-Bayesian Monte Carlo approximation method}\label{subsec3.3}

Having obtained the estimated parameters and estimated variance-covariance matrix of the mediator and outcome models, we propose a parameter-simulating quasi-Bayesian Monte Carlo approximation method to estimate and make inference on the causal effects. The Monte Carlo algorithm by \citet{medform} is extended to incorporate a functional mediator. To streamline notation, we denote the vector of estimated basis coefficients from the mediator model with $\widehat{\boldsymbol{\Omega}}_M = \widehat{\mathbf{b}}$ and let the estimated variance-covariance matrix for the estimated basis coefficients be $\widehat{\boldsymbol{\Sigma}}_M$. Likewise, the estimated parameters of the outcome model are denoted as $\widehat{\boldsymbol{\Omega}}_Y = \left(\widehat{\boldsymbol{\alpha}}^\top, \widehat{\boldsymbol{\gamma}}^\top\right)^\top$, and the estimated variance-covariance matrix for the estimated parameters is denoted as $\widehat{\boldsymbol{\Sigma}}_Y$. Specifically, we repeatedly simulate $Q$ sets of model parameters from the respective asymptotic sampling distributions of the mediator and outcome models, impute the potential outcomes to calculate the causal effects in each of the $Q$ simulated samples, and take the sample medians of the $Q$ copies of causal effects as point estimates. Algorithm \ref{alg:1} outlines the detailed procedures.
\begin{algorithm}[h]
\caption{Parameter-based Monte Carlo Simulation of the Causal Effects}
\begin{algorithmic}[1]
\Require $\widehat{\boldsymbol{\Omega}}_M$, $\widehat{\boldsymbol{\Sigma}}_M$, $\widehat{\boldsymbol{\Omega}}_Y$, $\widehat{\boldsymbol{\Sigma}}_Y$, $Q$, $n$
\Ensure $\widehat{\tau}_{\textsubscript{TE}}$, $\widehat{\tau}_{\textsubscript{NIE}}(a)$, $\widehat{\tau}_{\textsubscript{NDE}}(1-a)$, $\widehat{\sigma}_{\textsubscript{TE}}$, $\widehat{\sigma}_{\textsubscript{NIE$(a)$}}$, $\widehat{\sigma}_{\textsubscript{NDE$(1-a)$}}$, $\text{CI}_{95\%}(\tau_{\textsubscript{TE}})$, $\text{CI}_{95\%}(\tau_{\textsubscript{NIE}}(a))$, $\text{CI}_{95\%}(\tau_{\textsubscript{NDE}}(1-a))$
\State $q \gets 1$
\While{$q \leq Q$}
        \State $\boldsymbol{\Omega}_M^q \gets \mathcal{N}(\widehat{\boldsymbol{\Omega}}_M, \widehat{\boldsymbol{\Sigma}}_M)$
        \State $\boldsymbol{\Omega}_Y^q \gets \mathcal{N}(\widehat{\boldsymbol{\Omega}}_Y, \widehat{\boldsymbol{\Sigma}}_Y)$
        \State $i \gets 1$
        \While{$i \leq n$}
            \For{$a \in \{0, 1\}$}
                \State $\mathbf{M}_i(a) \gets \text{Mediator Model}(\boldsymbol{\Omega}_M^q) \mid A_i = a$
                \State $Y_i(a, \mathbf{M}_i(a)) \gets \text{Outcome Model}(\boldsymbol{\Omega}_Y^q) \mid A_i = a, \mathbf{M}_i = \mathbf{M}_i(a)$
                \State $Y_i(a, \mathbf{M}_i(1-a)) \gets \text{Outcome Model}(\boldsymbol{\Omega}_Y^q) \mid A_i = a, \mathbf{M}_i = \mathbf{M}_i(1-a)$
            \EndFor
            \State $i \gets i + 1$
        \EndWhile
    \State $\widehat{\tau}_{\textsubscript{TE}}^q \gets \operatorname{mean}(Y_i(1, \mathbf{M}_i(1)) - Y_i(0, \mathbf{M}_i(0)))$        
    \For{$a \in \{0, 1\}$}    
        \State $\widehat{\tau}_{\textsubscript{NIE}}^q(a) \gets \operatorname{mean}(Y_i(a, \mathbf{M}_i(a)) - Y_i(a, \mathbf{M}_i(1-a)))$
        \State $\widehat{\tau}_{\textsubscript{NDE}}^q(1-a) \gets \operatorname{mean}(Y_i(a, \mathbf{M}_i(1-a)) - Y_i(1-a, \mathbf{M}_i(1-a)))$
    \EndFor
    \State $q \gets q + 1$
\EndWhile
\State $\widehat{\tau}_{\textsubscript{TE}} \gets \operatorname{median}(\widehat{\tau}_{\textsubscript{TE}}^q)$
\State $\widehat{\tau}_{\textsubscript{NIE}}(a) \gets \operatorname{median}(\widehat{\tau}_{\textsubscript{NIE}}^q(a))$
\State $\widehat{\tau}_{\textsubscript{NDE}}(1-a) \gets \operatorname{median}(\widehat{\tau}_{\textsubscript{NDE}}^q(1-a))$
\State $\widehat{\sigma}_{\textsubscript{TE}} \gets \operatorname{sd}(\widehat{\tau}_{\textsubscript{TE}}^q)$
\State $\widehat{\sigma}_{\textsubscript{NIE$(a)$}} \gets \operatorname{sd}(\widehat{\tau}_{\textsubscript{NIE}}^q(a))$ 
\State $\widehat{\sigma}_{\textsubscript{NDE$(1-a)$}} \gets \operatorname{sd}(\widehat{\tau}_{\textsubscript{NDE}}^q(1-a))$
\State $\text{CI}_{95\%}(\tau_{\textsubscript{TE}}) \gets (\operatorname{quantile}(\widehat{\tau}_{\textsubscript{TE}}^q, 0.025), \operatorname{quantile}(\widehat{\tau}_{\textsubscript{TE}}^q, 0.975))$
\State $\text{CI}_{95\%}(\tau_{\textsubscript{NIE}}(a)) \gets (\operatorname{quantile}(\widehat{\tau}_{\textsubscript{NIE}}^q(a), 0.025), \operatorname{quantile}(\widehat{\tau}_{\textsubscript{NIE}}^q(a), 0.975))$
\State $\text{CI}_{95\%}(\tau_{\textsubscript{NDE}}(1-a)) \gets (\operatorname{quantile}(\widehat{\tau}_{\textsubscript{NDE}}^q(1-a), 0.025), \operatorname{quantile}(\widehat{\tau}_{\textsubscript{NDE}}^q(1-a), 0.975))$
\end{algorithmic}
\label{alg:1}
\end{algorithm}

\subsection{A Gaussian copula approach to sensitivity analysis} \label{sec3.4}
Identifying the natural direct and indirect effects within our potential outcomes framework requires assumptions outlined in Section \ref{subsec2.2}. Specifically, the second part of Assumption \ref{assump:conditionalindep} requires that the mediator behaves as if randomised, conditional on treatment and baseline covariates. As well perceived among researchers, \citet{medform} warns that such an assumption is ``strong", ``nonrefutable" and ``must be made with care", since unmeasured mediator–outcome confounding may persist despite treatment being conditionally randomised given observed covariates. A widely accepted approach to address this concern is to conduct a sensitivity analysis, which assesses how effect estimates would change under varying degrees of unmeasured mediator–outcome confounding \citep{medform, vanderweele2010bias, albert2011generalized, tchetgen2012semiparametric}.

We propose a sensitivity analysis that involves a functional mediator and a zero-inflated count outcome. To assess the potential impact of unmeasured mediator–outcome confounding, we introduce a latent sensitivity parameter $\rho$ that captures the residual dependence between the mediator and the outcome not explained by observed covariates. The central idea is to couple the counterfactual distributions of the mediator and outcome through a Gaussian copula framework, allowing us to flexibly model their joint distribution while varying the degree of latent correlation. By systematically varying this sensitivity parameter, we can evaluate how robust the estimated causal effects are to potential violations of the no-unmeasured-M-Y-confounding assumption.

The Gaussian copula method by \citet{albert2011generalized} and \citet{wangzero} is adopted to facilitate a joint distribution between continuous and discrete variables. Moreover, since the Gaussian copula method requires scalar inputs for constructing empirical cumulative distribution functions, we summarise the functional mediator with a scalar quantity by computing its area under the curve (AUC). The AUC provides a meaningful summary of the mediator trajectory and facilitates the empirical estimation of its distribution, thereby enabling the transformation of the mediator to a latent normal scale. Specifically, the count outcome $Y \in \left\{0, 1, \ldots, C_Y\right\}$ takes $C_Y + 1$ possible values, where $C_Y$ is the assumed upper limit for $Y$. For instance, in the context of hospital readmissions within 60 days following a CABG surgery, $C_Y$ may represent a clinically plausible maximum number of readmissions. We use the area under the curve $\bar{M} = \int_0^1 M(t) dt$ as a scalar summary of the mediator process. Notationally, $P_{Y\left(a^\prime, \mathbf{m}\right)}\left(y\right) = P\left\{Y\left(a^\prime, \mathbf{m}\right) \leq y\right\}$ and $P_{\bar{M}\left(a\right)}\left(\bar{m}\right) = P\left\{\bar{M}\left(a\right) \leq \bar{m}\right\}$ represent the empirical cumulative distribution functions of the potential outcomes and potential mediator AUC's from the fitted mediator and outcome models, respectively. The Gaussian copula method then posits two marginally standard normally distributed latent variables $U_1$ and $U_2$ that admit $U_1 = \Phi^{-1}\left\{P_{\bar{M}\left(a\right)}\left(\bar{m}\right)\right\}$ and $U_2 = \Phi^{-1}\left\{P_{Y\left(a^\prime, \mathbf{m}\right)}\left(y\right)\right\}$, where $\Phi$ is the standard normal cumulative distribution function. $U_1$ and $U_2$ are additionally assumed to jointly follow a bivariate normal distribution with correlation coefficient $\rho$, for all $a$, $a^\prime$ and $\mathbf{m}$. The correlation coefficient $\rho$ reflects the strength of unmeasured confounding between the mediator and the outcome. Namely, when $\rho = 0$, the assumption of no unmeasured mediator-outcome confounding holds, while deviations from zero indicate increasing levels of such confounding.

Similar to \citet{wangzero}, a Monte Carlo procedure is nested into the sensitivity analysis to compute the conditional probabilities of the count outcome $P\left\{Y\left(a^\prime, \mathbf{m}_i\right) = c, c = 1, \ldots, C_Y \mid \mathbf{M}(a) = \mathbf{m}_i\right\}$ and estimate the mean potential outcome over a large number of draws of the mediator process values. The detailed procedures of the proposed sensitivity analysis, conditional on a fixed set of model parameters and a value of $\rho$, are outlined as Algorithm \ref{alg:2}. The algorithm can be performed across a range of varying values of $\rho$ to assess the robustness of effect estimates to different degrees of unmeasured mediator-outcome confounding. One can also nest the sensitivity analysis into the parameter-simulating quasi-Bayesian Monte Carlo method described in Section \ref{subsec3.3} to obtain confidence intervals.
\begin{algorithm}[h]
\caption{Sensitivity analysis via Gaussian copula}
\begin{algorithmic}[1]
\Require $\widehat{\boldsymbol{\Omega}}_M$, $\widehat{\boldsymbol{\Sigma}}_M$, $\widehat{\boldsymbol{\Omega}}_Y$, $\widehat{\boldsymbol{\Sigma}}_Y$, $Q_s$, $\rho$, $n$, $C_Y$
\Ensure $\widehat{\tau}_{\textsubscript{NIE}}^{\rho}(a)$, $\widehat{\tau}_{\textsubscript{NDE}}^{\rho}(1-a)$

\For{$i = 1$ to $n$}
    \For{$a \in \{0, 1\}$}
        \State Sample $\mathbf{m}_i \sim \text{Mediator Model}(\widehat{\boldsymbol{\Omega}}_M) \mid A_i = a$
        \State $\bar{m}_i \gets \int_0^1 \mathbf{m}_i dt$
        \State $U_1 \gets \Phi^{-1}\left\{P_{\bar{M}(a)}(\bar{m}_i)\right\}$
    
        \For{$q = 1$ to $Q_s$}
            \State Sample $U_2^{q, a} \sim N(\rho U_1, 1 - \rho^2)$
            \State $u_2^{q, a} \gets \Phi(U_2^{q, a})$
            \For{$c = 1$ to $C_Y$}
                \If{$u_2^{q, a} \in \left( P_{Y(a, \mathbf{m}_i)}(c - 1), P_{Y(a, \mathbf{m}_i)}(c) \right]$}
                    \State $I_{i c q}^{a} \gets 1$
                \Else
                    \State $I_{i c q}^{a} \gets 0$
                \EndIf
            \EndFor
    
            \State Sample $U_2^{q, 1 - a} \sim N(\rho U_1, 1 - \rho^2)$
            \State $u_2^{q, 1 - a} \gets \Phi(U_2^{q, 1 - a})$
            \For{$c = 1$ to $C_Y$}
                \If{$u_2^{q, 1 - a} \in \left( P_{Y(1 - a, \mathbf{m}_i)}(c - 1), P_{Y(1 - a, \mathbf{m}_i)}(c) \right]$}
                    \State $I_{i c q}^{1 - a} \gets 1$
                \Else
                    \State $I_{i c q}^{1 - a} \gets 0$
                \EndIf
            \EndFor
        \EndFor
    
        \For{$c = 1$ to $C_Y$}
            \State $\widehat{P}\left\{ Y(a, \mathbf{m}_i) = c \mid \mathbf{M}(a) = \mathbf{m}_i \right\} \gets \frac{1}{Q_s} \sum_{q=1}^{Q_s} I_{i c q}^{a}$
            \State $\widehat{P}\left\{ Y(1 - a, \mathbf{m}_i) = c \mid \mathbf{M}(a) = \mathbf{m}_i \right\} \gets \frac{1}{Q_s} \sum_{q=1}^{Q_s} I_{i c q}^{1 - a}$
        \EndFor
    \EndFor
\EndFor

\For{$a \in \{0, 1\}$}
\State $\widehat{E}\left\{Y(a, \mathbf{M}(a))\right\} \gets \frac{1}{n} \sum_{i=1}^{n} \sum_{c=1}^{C_Y} c \cdot \widehat{P}\left\{ Y(a, \mathbf{m}_i) = c \mid \mathbf{M}(a) = \mathbf{m}_i \right\}$
\State $\widehat{E}\left\{Y(1 - a, \mathbf{M}(a))\right\} \gets \frac{1}{n} \sum_{i=1}^{n} \sum_{c=1}^{C_Y} c \cdot \widehat{P}\left\{ Y(1 - a, \mathbf{m}_i) = c \mid \mathbf{M}(a) = \mathbf{m}_i \right\}$
\State $\widehat{\tau}_{\textsubscript{NIE}}^{\rho}(a) \gets \widehat{E}\left\{Y(a, \mathbf{M}(a))\right\} - \widehat{E}\left\{Y(a, \mathbf{M}(1-a))\right\}$
\State $\widehat{\tau}_{\textsubscript{NDE}}^{\rho}(1-a) \gets \widehat{E}\left\{Y(a, \mathbf{M}(1-a))\right\} - \widehat{E}\left\{Y(1-a, \mathbf{M}(1-a))\right\}$
\EndFor
\end{algorithmic}
\label{alg:2}
\end{algorithm}

\section{Simulation study}\label{sec4}

In this section, we assess the performance of the proposed approach on simulated datasets. Procedures for generating data are presented in Section \ref{subsec4.1}, followed by a discussion of empirical results from the simulation study, as well as comparisons of the proposed methods to an existing one in Section \ref{subsec4.2}.

\subsection{Data generation}\label{subsec4.1}

\begin{figure}[!t]
\centering
\includegraphics[scale=0.45]{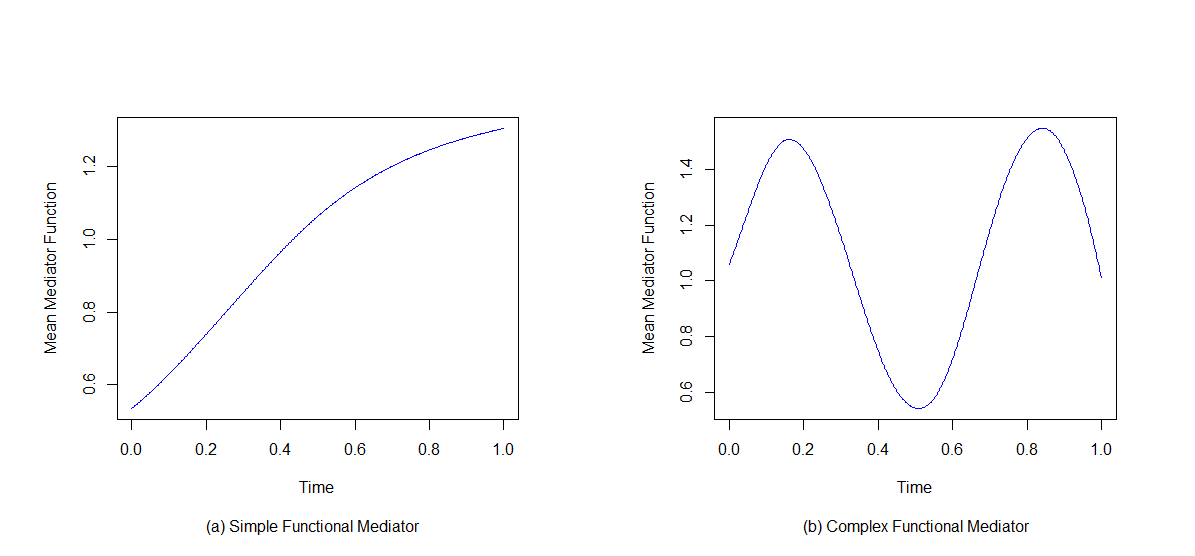}
\caption{Mean functions of the functional mediator from a $n = 1000$ generated sample: a simple functional mediator (left panel), a complex functional mediator (right panel).}
\label{fig:simpvscomp}
\end{figure}

We consider the time interval $\mathcal{T} = [0, 1]$ and evaluate two sample sizes, $n = 100$ and $n = 1000$. To evaluate the performance of the proposed methods in handling infrequent and irregular mediator observations, we generate $T_i \overset{\mathrm{iid}}{\sim} \operatorname{Poisson}(25)$ as the number of observations per subject to reflect the level of sparsity observed in our MIMIC-IV dataset. The discrete observations $\Tilde{M}(t_{ij})$ are assumed to randomly happen at $T_i$ time points on $[0, 1]$ and may be contaminated by measurement errors. Additionally, we explore two different forms of the underlying mediator process $M(t)$ as shown in Figure \ref{fig:simpvscomp}, resulting in a total of four scenarios. For each scenario, we generate $R = 1000$ replicated datasets corresponding to the specified sample size ($n = 100$ or $n = 1000$).

In each replication dataset, the binary treatment $A_i$ is assigned randomly with a probability of $0.5$ to either the treatment or control group. The baseline covariate $X_i$ is drawn from $N(1, 3)$. The underlying mediator process is generated over the time interval $\mathcal{T} = [0, 1]$ as
\begin{equation*} 
        M_{i}(t) = \beta_{0}(t) + \beta_{1}(t)A_i + \beta_{2}(t) X_i + \varepsilon_{i}(t), \quad i = 1, \ldots, n.
\end{equation*}
For scenarios with a simple functional mediator, $\beta_{0}(t) = 0$, $\beta_{1}(t) = 1.5\sin(0.5\pi t)$, $\beta_{2}(t) = 0.5$. The error term $\varepsilon_{i}(t)$ is generated from a multivariate normal distribution with mean zero and covariance $\Sigma(s, t)$, where $\Sigma(s, t) = \exp(-3|s - t|)$.  For scenarios with a complex functional mediator, $\beta_{0}(t) = 0$, $\beta_{1}(t) = \sin(3 \pi t) + 1$, $\beta_{2}(t) = 0.5$. The error term $\varepsilon_{i}(t)$ is generated from a multivariate normal distribution with mean zero and covariance $\Sigma(s, t)$, where $\Sigma(s, t) = 2\exp(-3|s - t|)$. The parameter values are chosen so that the true effect sizes are close among scenarios. Figure \ref{fig:simpvscomp} illustrates the shapes of a simple functional mediator and a complex one, where the simple mediator function is monotonically increasing on the time interval and the complex mediator function fluctuates as time progresses. The error-contaminated observations $\Tilde{M}(t_{ij})$ are then obtained by
\begin{equation*} 
        \Tilde{M}(t_{ij}) = M_{i}(t_{ij}) + \epsilon_{ij}, \quad i = 1, \ldots, n,\ j = 1, \ldots, T_i,\ \epsilon_{ij} \overset{\mathrm{iid}}{\sim} N(0, 0.1).
\end{equation*}

The end-point zero-inflated count outcome $Y_i$ is generated as in (\ref{eq:outcomemodel}), (\ref{eq:logisticlink}) and (\ref{eq:loglink}). For scenarios with a simple functional mediator, $\alpha_{0} = -3$, $\alpha_{1}(t) = -4(t - 0.5)^2 + 1$, $\alpha_{2} = 0.5$, $\alpha_{3} = 0.5$, and $\gamma_0 = 1$, $\gamma_1(t) = 0.5$, $\gamma_2 = 1$, $\gamma_3 = -0.1$. For scenarios with a complex functional mediator, $\alpha_{0} = -3.5$, $\alpha_{1}(t) = 0.07\sin(3 \pi t) + 1$, $\alpha_{2} = 0.5$, $\alpha_{3} = 0.5$, and $\gamma_0 = 1$, $\gamma_1(t) = 0.5$, $\gamma_2 = 1$, $\gamma_3 = -0.1$. The parameter values are chosen so that there are approximately $35\%$ of the count outcomes in each dataset are zero and the true effect sizes are close among scenarios.

The true causal direct, indirect and total effects for each scenario are empirically computed by generating a dataset containing $100,000$ subjects. 

\subsection{Empirical results}\label{subsec4.2}

\begin{table}[t]
\caption{Comparison of performance under different sample sizes $n$ with a simple functional mediator: error-contaminated observations happen randomly at $T_i \overset{\mathrm{iid}}{\sim} \operatorname{Poisson}(25)$ time points on $[0,1]$, the number of replication datasets $R = 1000$, the basis system used for the mediator model is cubic splines with $nbasis = 20$, the number of functional principal components used for the outcome model $\widehat{K} = 2$, the number of simulated sets of model parameters in Algorithm \ref{alg:1} $Q = 1000$.  \label{tabsimplesimulation}}
\begin{tabular*}{\textwidth}{@{\extracolsep{\fill}}lccccc@{\extracolsep{\fill}}}
\toprule%
Estimand & True Value  & BIAS (\%) & ASE & ESE & ECP (\%) \\
\midrule
\multicolumn{6}{@{}c@{}}{$n = 1000$} \\
\midrule
$\tau_{\textsubscript{TE}}$ & $5.28$ & $1.54$ & $0.410$ & $0.396$ & $94.4$ \\
$\tau_{\textsubscript{NIE}}(1)$ & $2.02$ & $2.13$ & $0.275$ & $0.278$ & $93.3$ \\
$\tau_{\textsubscript{NDE}}(0)$ & $3.26$ & $1.44$ & $0.327$ & $0.335$ & $93.5$ \\
$\tau_{\textsubscript{NIE}}(0)$ & $0.90$ & $0.31$ & $0.132$ & $0.138$ & $93.3$ \\
$\tau_{\textsubscript{NDE}}(1)$ & $4.38$ & $2.07$ & $0.425$ & $0.414$ & $94.0$ \\
\midrule
\multicolumn{6}{@{}c@{}}{$n = 100$} \\
\midrule
$\tau_{\textsubscript{TE}}$ & $5.28$ & $11.64$ & $2.006$ & $1.347$ & $95.6$ \\
$\tau_{\textsubscript{NIE}}(1)$ & $2.02$ & $23.26$ & $1.497$ & $1.031$ & $93.2$ \\
$\tau_{\textsubscript{NDE}}(0)$ & $3.26$ & $5.54$ & $1.127$ & $1.037$ & $95.5$ \\
$\tau_{\textsubscript{NIE}}(0)$ & $0.90$ & $15.31$ & $0.636$ & $0.479 $ & $95 .5$ \\
$\tau_{\textsubscript{NDE}}(1)$ & $4.38$ & $11.73$ & $1.772$ & $1.301$ & $96.0$ \\
\midrule
\end{tabular*}
\end{table}

\begin{table}[t]
\caption{Comparison of performance under different sample sizes $n$ with a complex functional mediator: error-contaminated observations happen randomly at $T_i \overset{\mathrm{iid}}{\sim} \operatorname{Poisson}(25)$ time points on $[0,1]$, the number of replication datasets $R = 1000$, the basis system used for the mediator model is cubic splines with $nbasis = 20$, the number of functional principal components used for the outcome model $\widehat{K} = 2$, the number of simulated sets of model parameters in Algorithm \ref{alg:1} $Q = 1000$. \label{tabcomplexsimulation}}
\begin{tabular*}{\textwidth}{@{\extracolsep{\fill}}lccccc@{\extracolsep{\fill}}}
\toprule%
Estimand & True value  & BIAS (\%) & ASE & ESE & ECP (\%) \\
\midrule
\multicolumn{6}{@{}c@{}}{$n = 1000$} \\
\midrule
$\tau_{\textsubscript{TE}}$ & $5.20$ & $2.57$ & $0.338$ & $0.349$ & $90.9$ \\
$\tau_{\textsubscript{NIE}}(1)$ & $1.93$ & $8.82$ & $0.240$ & $0.262$ & $85.4$ \\
$\tau_{\textsubscript{NDE}}(0)$ & $3.27$ & $1.14$ & $0.269$ & $0.268$ & $94.3$ \\
$\tau_{\textsubscript{NIE}}(0)$ & $0.88$ & $7.46$ & $0.131$ & $0.135$ & $91.3$ \\
$\tau_{\textsubscript{NDE}}(1)$ & $4.32$ & $1.61$ & $0.365$ & $0.364$ & $94.5$ \\
\midrule
\multicolumn{6}{@{}c@{}}{$n = 100$} \\
\midrule
$\tau_{\textsubscript{TE}}$ & $5.20$ & $12.79$ & $1.628$ & $1.35$ & $92.7$ \\
$\tau_{\textsubscript{NIE}}(1)$ & $1.93$ & $30.89$ & $1.260$ & $1.005$ & $91.1$ \\
$\tau_{\textsubscript{NDE}}(0)$ & $3.27$ & $2.11$ & $0.937$ & $0.929$ & $94.6$ \\
$\tau_{\textsubscript{NIE}}(0)$ & $0.88$ & $25.12$ & $0.579$ & $0.493$ & $94.2$ \\
$\tau_{\textsubscript{NDE}}(1)$ & $4.32$ & $10.86$ & $1.499$ & $1.326$ & $94.4$ \\
\midrule
\end{tabular*}
\end{table}
For implementation, we specify $K_b = 20$ cubic spline basis functions for the mediator model to provide a flexible representation of the coefficient functions prior to penalisation. In the outcome model, we incorporate the first $\widehat{K}$ FPCs of the mediator process, where $\widehat{K}$ is data-driven and determined to ensure at least 90\% of the total variance is accounted for. Consequently, the PACE procedure selects $\widehat{K} = 2$ for both the simple and complex mediator processes. Subsequently, from the Monte Carlo approximation algorithm, we obtain a point estimate, a sample standard deviation and a $95\%$ confidence interval for each of the causal direct, indirect and total effects for each of the $R = 1000$ replication datasets. We use the total effect as an example to illustrate the calculation of the percent absolute bias (BIAS), empirical coverage probability (ECP), average standard errors (ASE), and empirical standard errors (ESE). For the $r$-th replication dataset, where $r = 1, \ldots, R$, we denote the point estimate as $\widehat{\tau}^r_{\textsubscript{TE}}$, the sample standard deviation as $\widehat{\sigma}^r_{\textsubscript{TE}}$, and the $95\%$ confidence interval as $\text{CI}_{95\%}^r(\tau_{\textsubscript{TE}})$. The performance measures are calculated as follows,
\begin{align*}
\text{BIAS} &= \frac{|\sum_{r=1}^R\widehat{\tau}^r_{\textsubscript{TE}} / R - \tau_{\textsubscript{TE}}|}{|\tau_{\textsubscript{TE}}|} \times 100\%,\\
\text{ECP} &= \frac{\sum_{r = 1}^R\mathbbm{1}(\tau_{\textsubscript{TE}} \in \text{CI}_{95\%}^r(\tau_{\textsubscript{TE}}))}{R} \times 100\%,\\
\text{ASE} &= \frac{\sum_{r = 1}^{R}\widehat{\sigma}^r_{\textsubscript{TE}}}{R},\\
\text{ESE} &= \sqrt{\frac{\sum_{r=1}^{R}\left(\widehat{\tau}^r_{\textsubscript{TE}}-\sum_{r=1}^R\widehat{\tau}^r_{\textsubscript{TE}}/ R\right)^2}{R-1}},
\end{align*}
where $\tau_{\textsubscript{TE}}$ refers to the true total effect.

Tables \ref{tabsimplesimulation} and \ref{tabcomplexsimulation} summarise the empirical results from the simulation studies. It can be observed that the proposed method performs well for large samples. This is consistent with established findings in the literature, which emphasise that for infrequently and irregularly observed functional data, reliable estimation requires a sufficiently large sample size so that the pooled observation times across subjects approximate a dense coverage of the domain \citep{yao2005sparse, zhang2016from, nie2022recovering}. With a large sample size, the method yields little bias when estimating the causal effects. As expected, small sample sizes are associated with increased variability in effect estimates. The empirical coverage probabilities are close to the nominal $95\%$. The lack of aggregated coverage of the domain as a result of small samples presents itself as the most important factor affecting the performance of the proposed method in estimating the causal effects. We observe a general pattern of dramatically increased bias and variability in scenarios with a small sample compared to a large one. Despite not having as notable impacts on the bias and variability of the effect estimates, a more complex functional form of the mediator results in larger bias, as expected. 

Overall, the proposed method shows solid performance across the scenarios with a large sample size. This suggests that the method is particularly well suited for large-scale observational studies, where the sample size is sufficient to support reliable estimation and inference involving infrequently and irregularly observed functional data. An example of such a setting is our application to the MIMIC-IV dataset, which includes nearly 5000 individuals.

Although our proposed methods are designed to accommodate sparse functional mediators, they can also be applied in settings where the mediator is densely observed. We note, however, that when the functional mediator is dense, more efficient estimation strategies are available in the functional data analysis literature and could be considered. To benchmark performance, we conduct additional simulation studies where the functional mediator is densely observed without measurement error, and compare the proposed methods to an existing \texttt{R} package \texttt{funmediation} \citep{funmediationpack}, which also performs functional causal mediation analysis for an end-point outcome (see Section A in the Web-based supporting materials for full details). Specifically, methods in \texttt{funmediation} assume confounding variables and a binary treatment at baseline, a functional mediator, and an end-point continuous or binary outcome, by adopting linear functional structural equation models similar to those in \citet{lindquist12}. Obviously, \texttt{funmediation} is not intended for our settings where the outcome is a zero-inflated count. First, the outcome scalar-on-function regression models used by \texttt{funmediation} assume either a Gaussian distribution or a Bernoulli distribution for the outcome, which is inadequate for count data, especially one with excess zeros. Second, the linear functional structural equation models imply the equivalence between $\tau_{\textsubscript{NIE}}(1)$ and $\tau_{\textsubscript{NIE}}(0)$, and thus also between $\tau_{\textsubscript{NDE}}(0)$ and $\tau_{\textsubscript{NDE}}(1)$. However, this is not the case when nonlinearity is present between the treatment, mediator, and outcome, which means the use of linear models cannot capture this in our settings. Consequently, while \texttt{funmediation} performs reasonably close to the proposed methods when estimating the total effect, it fails to reliably estimate the natural direct and indirect effects across all data-generating mechanisms. Therefore, the proposed methods offer a more suitable approach for reliably estimating causal effects in the context of a functional mediator and a zero-inflated count outcome. By addressing the limitations of linear models and capturing the nonlinearity between treatment, mediator, and outcome, the methods provide an effective framework for distinguishing between natural direct and indirect effects across different treatment statuses in causal mediation analysis. 

\section{Gender, central venous pressure, and rehospitalisations after CABG in MIMIC-IV}\label{sec5}

Our research question in this paper is two-fold. Firstly, we would like to investigate the causal effect of gender on the number of hospital readmissions within 60 days of discharge from the primary hospital stay among patients who underwent CABG in the MIMIC-IV database. Secondly, we intend to examine whether the time-varying CVP measured over the first 24 hours of post-surgical ICU stay mediates the causal pathway from gender to rehospitalisations. Consequently, we arrive at a functional causal mediation analysis with gender being a binary treatment, time-varying CVP during the postoperative ICU stay being a functional mediator, and the 60-day readmission count as a zero-inflated count outcome. Admittedly, the use of non-manipulable characteristics, such as gender or race, as treatment variables in causal inference has generated considerable debate. We provide a discussion and justification of the choices in this paper in Section \ref{subsec5.1}. The procedures for extracting data from MIMIC-IV, as well as the research question in the context of the final study data, are introduced in Section \ref{subsec5.2}. Section \ref{subsec5.3} provides a discussion of the findings and insights.

\subsection{Gender as a treatment in a causal framework to study rehospitalisations after CABG}\label{subsec5.1}
\begin{figure}[!t]
\centering
\begin{tikzpicture}[node distance=4cm]
\node[box] (gender) {Gender};
\node[box, right of=gender] (cabg) {CABG};
\node[box, right of=cabg] (cvp) {Time-varying CVP};
\node[box, right of=cvp] (readmit) {Rehospitalisations};
\node[box, above of=gender, yshift=-2cm] (baseline) {Baseline Covariates};

\draw[path] (gender) -- (cabg);
\draw[path] (cabg) -- (cvp);
\draw[path] (cvp) -- (readmit);
\draw[path] ($ (gender.south) + (-3mm, 0) $) |- ([yshift=-12mm]gender.south) -| ([xshift=3mm]readmit.south);
\draw[path] (cabg)  |- ([yshift=-6mm]cabg.south)  -| ([xshift=-3mm]readmit.south);
\draw[path] 
  ($ (gender.south) + (3mm, 0) $) 
  -- ($ (gender.south) + (3mm, -9mm) $)
  -| (cvp);
\draw[path] (baseline) -- (cabg);
\draw[path] (baseline) -- (cvp);
\draw[path] (baseline) -- (readmit);

\end{tikzpicture}
\caption{A causal diagram for conceptualising the assumed causal structure for an imaginary broader population.}
\label{fig:cabgdag}
\end{figure}

\begin{figure}[!t]
\centering
\begin{tikzpicture}[node distance=4cm]
\node[box] (gender) {Gender};
\node[box, right of=gender] (cvp) {Time-varying CVP};
\node[box, right of=cvp] (readmit) {Rehospitalisations};
\node[box, above of=gender, yshift=-2cm] (baseline) {Baseline Covariates};

\draw[path] (gender) -- (cvp);
\draw[path] (cvp) -- (readmit);
\draw[path] (gender) |- ([yshift=-6mm]gender.south) -| (readmit);
\draw[path] (baseline) -- (readmit);
\draw[path] (baseline) -- (cvp);
\draw[dotted,<->] (baseline) to (gender);

\end{tikzpicture}
\caption{The causal diagram under analysis after conditioning on CABG.}
\label{fig:analysisdag}
\end{figure}
Before presenting the data structure and empirical results, we begin by addressing a key conceptual consideration in our study: the treatment of gender as a causal exposure. The inclusion of non-manipulable characteristics, such as gender or race, as treatment variables in causal inference has prompted considerable philosophical and methodological debate.

A prominent line of critique maintains that causal estimands must be grounded in well-specified hypothetical interventions, a view encapsulated by the doctrine of “no causation without manipulation” introduced by \citet{holland1986statistics}. \citet{hernan2004definition} expands on this perspective by formalising causal effects within the potential outcomes framework and emphasising that meaningful causal inference requires interventions that are clearly defined and, ideally, implementable in practice. Without such clarity, causal questions and their answers risk becoming ill-posed or ambiguous. Illustrating this concern, \citet{hernan2008does} examines the case of obesity, an exposure that cannot be reduced to a single intervention. They argue that unless a causal question specifies whether obesity is intervened upon through diet, medication, or surgery, it is unclear which potential outcomes are being contrasted, making the estimand difficult to interpret. In a related critique, \citet{kaufman2008epidemiologic} highlights the challenges of treating race and ethnicity as causal exposures, cautioning that such analyses may be misleading if they fail to specify which dimension of these constructs is being considered, such as phenotype, social classification, or lived experience. \citet{didden2025targeting} offers a comprehensive summary of these critiques and proposes an alternative approach based on interventional effects. Rather than relying on potential outcomes that assume interventions on non-manipulable characteristics such as gender, this framework focuses on hypothetical interventions on modifiable mediators. It estimates how much of the observed disparity would remain if the mediator distribution were shifted to that observed under a different exposure level. This approach circumvents the conceptual dilemma of defining causal effects while preserving the policy relevance of mediation analysis.

In contrast to the critiques above, several researchers have defended the use of causal estimands involving non-manipulable exposures such as gender or race, provided the underlying assumptions and interpretations are made explicit. \citet{vanderweele2012natural} proposes that causal effects involving non-manipulable exposures can still be meaningfully conceptualised if we imagine a function that captures the full set of inputs governing the outcome, analogous to a natural law. An ``all-cause" model in the form of $Y = f\left(Gender, Age, Race, Comorbidities, Health, \ldots \right)$, where potential outcomes can be generated by varying, for example, gender, while holding all other components constant. Although it becomes immediately obvious that such functions are rarely known in practice, the authors argue that this framework allows for scientifically coherent counterfactual reasoning, even in the absence of manipulability. This ``all causes” model parallels the logic of physical sciences, where contrary-to-fact predictions can be derived from system dynamics rather than manipulations. While it is almost never possible to fully characterise the ``all-causes” function that governs an outcome, estimation of causal effects is still feasible in settings with randomisation or, more generally, under the assumption of conditional independence. In the latter case, adjustment for a set of covariates $\mathbf{X}$ that is sufficient to block all backdoor paths between the exposure and outcome allows for identification, even if $\mathbf{X}$ does not coincide exactly with the full set of inputs to the ``all-causes” model. \citet{vanderweele2014causal} adopts a cautious position, proposing that causal interpretations of race can still be made if race is conceptualised as a composite variable encompassing social, cultural, and biological attributes. Within this framework, they argue that adjusting for mediators such as early-life and adult socioeconomic status enables the decomposition of observed disparities into modifiable pathways, thereby permitting meaningful causal interpretation even without direct interventions on race itself. \citet{glymour2014commentary} goes even further by rejecting the premise that manipulation is necessary for a variable to be considered a cause. They argue that race and sex can be causes without being reduced to manipulable components, and that causal claims may be formed on systematic dependence rather than hypothetical interventions. The commentary also highlights that many variables routinely treated as exposures in observational studies, such as education, occupation, social circle or weight, are also difficult to intervene upon. Yet they rarely face similar controversies. Instead, they propose an aetiological notion of causation that differs from the interventionist perspective, which may be more appropriate for identifying explanatory factors. Using examples such as sex-based discrimination in employment, they contend that causal reasoning remains meaningful even when interventions are infeasible or conceptually awkward.

In light of this ongoing debate, we acknowledge that recognising gender as a causal treatment in studying rehospitalisations after CABG requires careful conceptualisation. Following the perspective that gender may plausibly be treated as a cause in the sense discussed by \citet{vanderweele2012natural} and \citet{glymour2014commentary}, we aim to explore gender as a treatment variable where the systematic differences in physiology, clinical management, or aftercare may lead to downstream effects. Specifically, we posit a causal paradigm for a broader population, not restricted to those who have undergone the surgery, in Figure \ref{fig:cabgdag}. Here, gender and baseline covariates, such as age and race, independently cause the receipt of CABG surgery. CABG, in turn, affects CVP, a time-varying physiological parameter measured during the first 24 hours of post-operative ICU stay. CVP then influences the number of hospital readmissions within 60 days of discharge, making it a potential mediator of the effect of gender on the outcome. The baseline covariates may act as mediator-outcome confounders. For this imaginary broader population, one important nuance is that CABG acts as a collider between gender and baseline covariates. When the target population of our study is restricted to patients who received the surgery, effectively conditioning on the receipt of CABG, a backdoor path is opened between gender and baseline covariates, as illustrated by the dotted double-headed arrow in Figure \ref{fig:analysisdag}. Such induced dependence renders gender a non-randomised treatment in the study. As a result, adjustment for baseline covariates is required to obtain unbiased estimates of the effects of gender.

\subsection{Study data from MIMIC-IV}\label{subsec5.2}

For this study, all data were retrieved from version 2.2 of the Medical Information Mart for Intensive Care IV (MIMIC-IV) database. MIMIC-IV is a large publicly available electronic health record (EHR) database containing 299,712 patients, 431,231 hospital admissions, and 73,181 ICU stays. The database, sourced from an EHR system of the hospital and a clinical information system of the ICU, includes all deidentified medical records corresponding to patients admitted to an ICU or the emergency department of the Beth Israel Deaconess Medical Center (BIDMC) between 2008 and 2019. 

\begin{figure}[!t]
\centering
\begin{tikzpicture}[node distance=2cm and 2cm]

    \node (start) [block] {Patients in MIMIC IV\\ n=299,712};
    \node (step1) [block, below of=start] {5,655 patients included};
    \node (step2) [block, below of=step1] {5,080 patients included};
    \node (step3) [block, below of=step2] {5,012 patients included};
    \node (final) [block, below of=step3] {Selected cohort\\ n=4,961};

    \node (ex1) [block, right of=step1, xshift=3cm, yshift=1cm] {Patients without coronary artery bypass grafting\\ n=294,057};
    \node (ex2) [block, right of=step2, xshift=3cm, yshift=1cm] {Patients transferred to ICU more than 1 day after surgery or not at all\\ n=575};
    \node (ex3) [block, right of=step3, xshift=3cm, yshift=1cm] {Patients died at primary hospital admission\\ n=68};
    \node (ex4) [block, right of=final, xshift=3cm, yshift=1cm] {Patients without first-24-hour CVP data\\ n=51};

    \draw [arrow] (start) -- (step1);
    \draw [arrow] (step1) -- (step2);
    \draw [arrow] (step2) -- (step3);
    \draw [arrow] (step3) -- (final);

    \draw [arrow] (start.south) ++(0,-0.5) -- ++(0cm,0) |- (ex1.west);
    \draw [arrow] (step1.south) ++(0,-0.5) -- ++(0cm,0) |- (ex2.west);
    \draw [arrow] (step2.south) ++(0,-0.5) -- ++(0cm,0) |- (ex3.west);
    \draw [arrow] (step3.south) ++(0,-0.5) -- ++(0cm,0) |- (ex4.west);

\end{tikzpicture}
\caption{A flow chart of cohort selection procedures from MIMIC-IV.}
\label{fig:cohortselection}
\end{figure}

A total of 5,655 patients in MIMIC-IV who underwent CABG were identified using the Ninth and Tenth Revisions of the International Classification of Diseases (ICD-9 code 361 and ICD-10 code 021). A full list of collected variables, including general demographics, clinical characteristics, and time-varying measurements, is provided in Section B of the Web-based supporting materials. The final cohort was formed using the following exclusion criteria: (a) patients who were not transferred to the ICU during the primary CABG hospital admission or were transferred more than one day after surgery; (b) patients who died during the primary CABG admission; and (c) patients without CVP measurements during the first 24 hours of their postoperative ICU stay. As shown in Figure \ref{fig:cohortselection}, 4,961 patients were retained for analysis.
\begin{figure}[!t]
\centering
\includegraphics[width=0.95\linewidth]{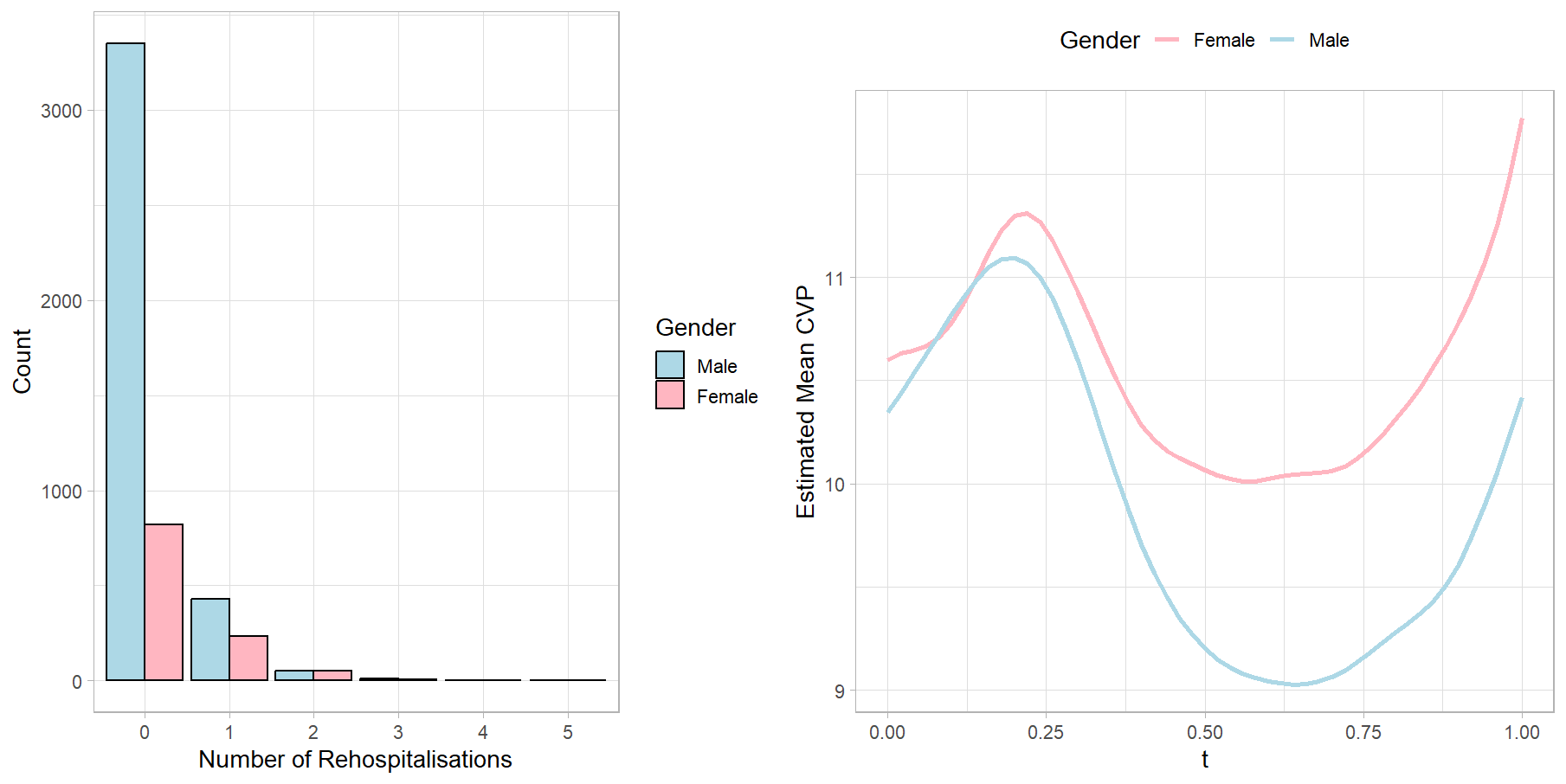}
\caption{Left panel: a comparison of distributions of 60-day rehospitalisations after CABG between genders. Right panel: a comparison of estimated mean CVP functions during the first 24 hours of post-surgical ICU stay between genders.}
\label{fig:readmissiondistandcvpcurveMvsF}
\end{figure}

\begin{figure}[!t]
\centering
\includegraphics[width=0.95\linewidth]{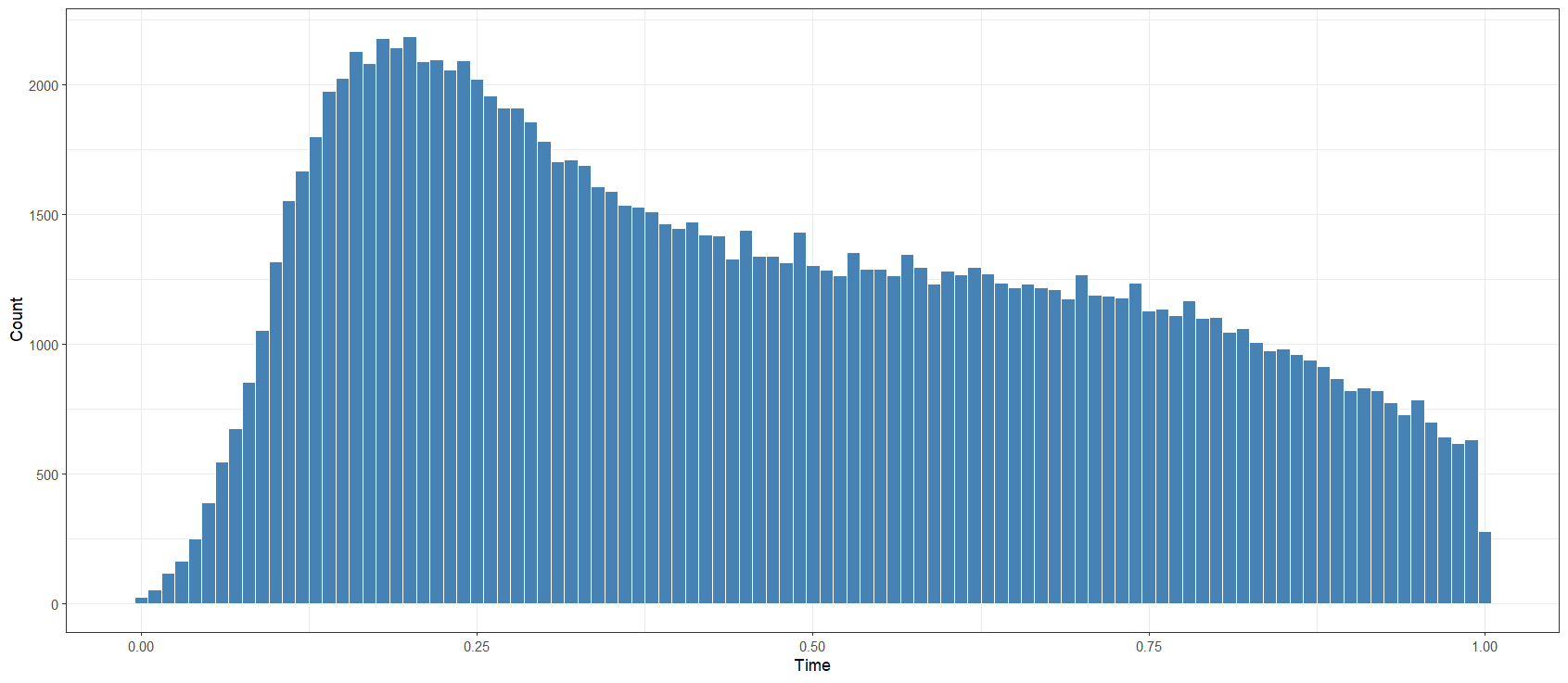}
\caption{Distribution of discrete CVP measurement times in the study data: the x-axis represents the rescaled time.}
\label{fig:obstime}
\end{figure}

The left panel of Figure \ref{fig:readmissiondistandcvpcurveMvsF} reveals pronounced concentration at zero in the distribution of 60-day hospital readmissions following CABG surgery. Remarkably, 4,170 out of 4,961 patients, approximately $84\%$, did not require any readmission during this period, underscoring the importance of employing modelling approaches that appropriately account for zero-inflation in the outcome. The disparity between genders is notable, with $87\%$ of male patients experiencing no readmissions within 60 days, compared to only $74\%$ of female patients. This motivates a causal analysis of whether there is a causal effect of gender on 60-day rehospitalisation counts. We use curves in the right panel of Figure \ref{fig:readmissiondistandcvpcurveMvsF} to reveal the difference in the estimated mean CVP functions during the first 24 hours of postoperative ICU stay between male and female patients. While women have a nearly consistently higher CVP than men on average, the shapes of the trajectories are intriguing. The mean CVP functions for males and females exhibit a characteristic pattern: an initial surge, followed by a gradual decline, and a subsequent secondary increase. Along with evidence of CVP's prognostic role for post-CABG complications, we aim to examine the mediating effect of not only the baseline CVP measurement but also the entire CVP process. As shown in Figure \ref{fig:obstime}, the CVP measurements are infrequently and irregularly distributed over the 24-hour window, with the mean number of observations per patient being around $25$. This confirms the need for the proposed methods for a sparse longitudinal mediator. For subsequent analysis, we restrict adjustment to baseline age and race, as these are thought not to be causally influenced by gender. On the other hand, comorbidities and postoperative physiological measurements, such as various severity scores, may be downstream consequences of gender and are thus more appropriately treated as mediators rather than confounders. This decision aligns with recommendations from existing literature, particularly in disparities research, which cautions against controlling for post-treatment variables when estimating the effect of a fixed characteristic. \citet{vanderweele2014causal} emphasises that adjusting for mediators such as adult socioeconomic status in race effect analyses partitions the total effect and distorts the estimated disparity attributable to the treatment. Similarly, \citet{knox2020administrative} highlights how conditioning on variables affected by race, such as arrest rates, can induce post-treatment bias and lead to invalid causal conclusions. To preserve the interpretability of our estimated effect of gender, we therefore refrain from including variables plausibly influenced by it. The treatment-mediator interaction is tested but also excluded due to a lack of statistical significance. As much as we aim to avoid adjusting for variables that may lie on the causal pathway from gender to outcome, we acknowledge that including only two baseline covariates may raise concerns about the validity of the conditional independence assumption due to potential unmeasured confounding. To address this, we apply the proposed sensitivity analysis procedures introduced in Section \ref{sec3.4}, the results of which are presented in the following section.


\subsection{Findings and insights}\label{subsec5.3}

We posit the causal framework described in Section \ref{sec2} and apply the proposed methods from Section \ref{sec3} to investigate the causal relationship between gender, time-varying CVP and 60-day hospital readmissions after CABG. Section \ref{subsubsec5.3.1} presents findings from functional data analysis related to time-varying CVP. We discuss the results and insights provided by causal mediation analysis and sensitivity analysis in \ref{subsubsec5.3.2}.

\subsubsection{Results from functional data analysis}\label{subsubsec5.3.1}

\begin{figure}[!t]
    \centering
    \includegraphics[width=0.95\linewidth]{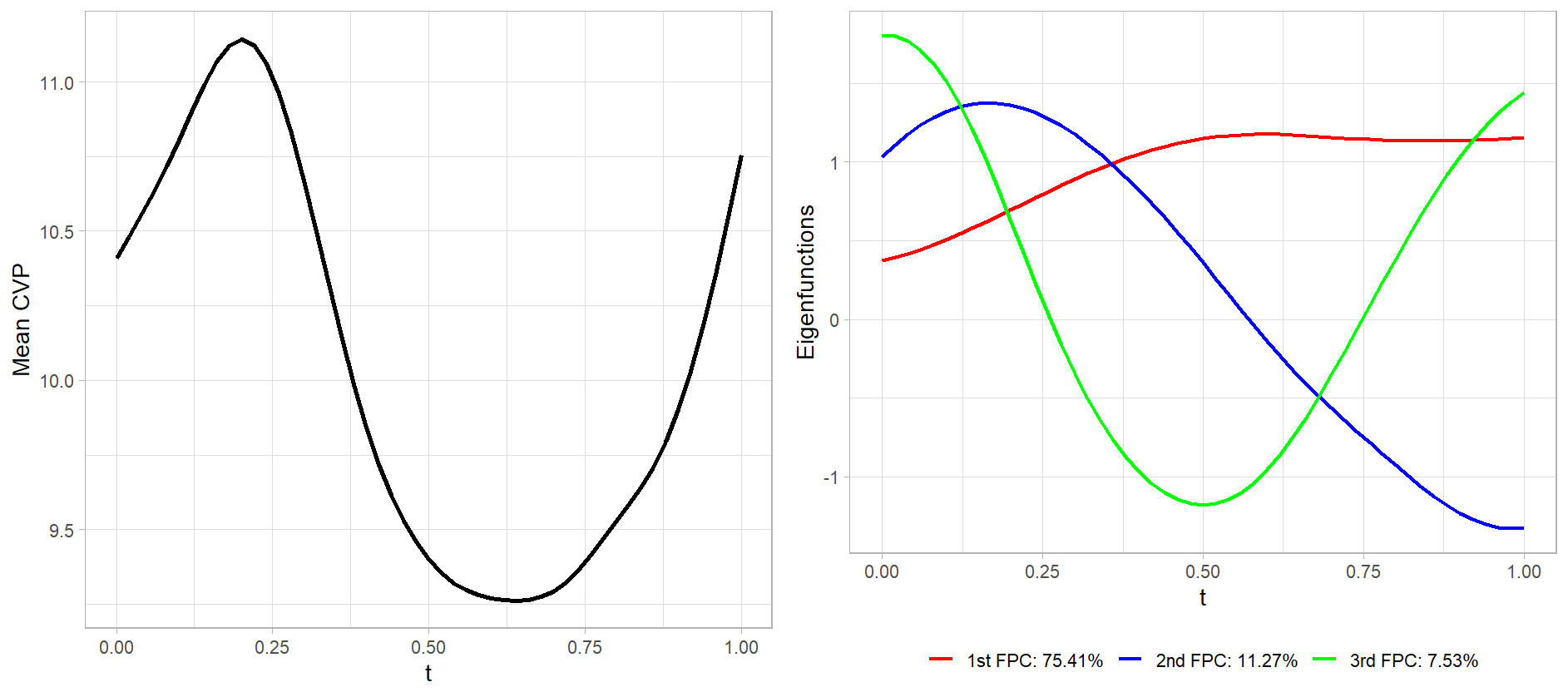}
    \caption{Left panel: the overall mean function of the mediator, that is, the CVP function during the first 24 hours of the post-surgical ICU stay. Right panel: the first three FPCs of the CVP function.}
    \label{fig:meanCVPandFPCS}
\end{figure}

\begin{figure}[!t]
    \centering
    \includegraphics[width=0.95\linewidth]{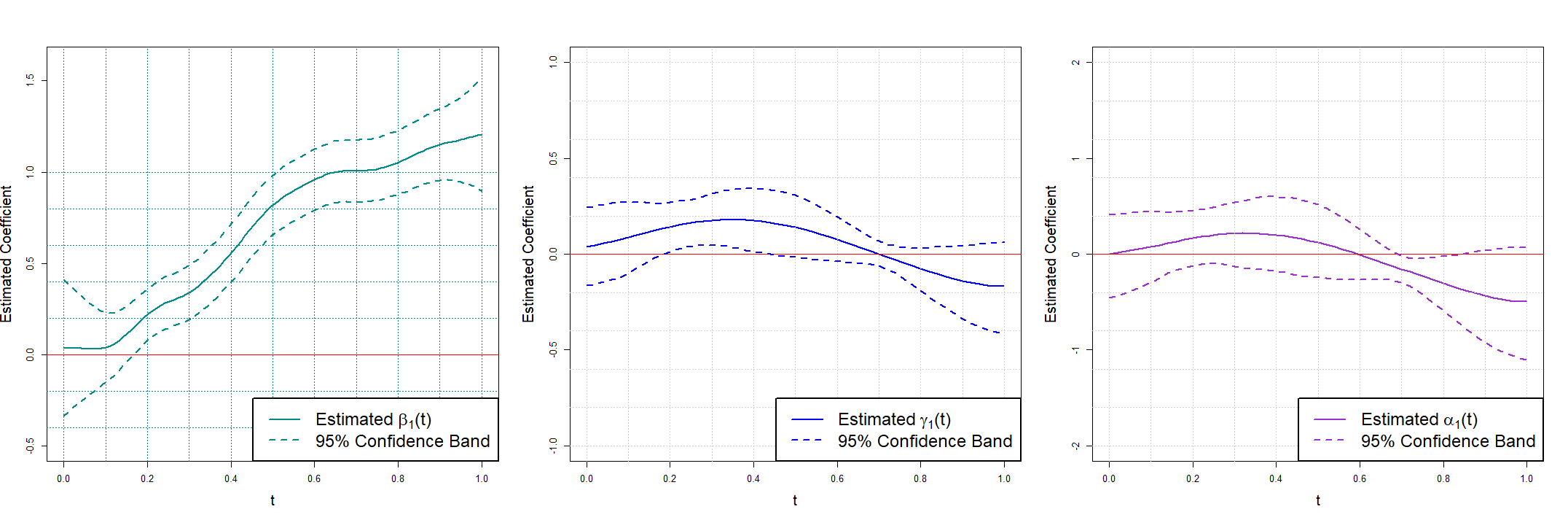}
    \caption{Estimated coefficient functions from the mediator and outcome models: the estimated coefficient function for the treatment gender in the mediator model $\widehat{\beta}_1(t)$ (left panel); the estimated coefficient function for the mediator CVP function in the outcome model count component $\widehat{\gamma}_1(t)$ (middle panel); the estimated coefficient function for the mediator CVP function in the outcome model zero component $\widehat{\alpha}_1(t)$.}
    \label{fig:estimatedcoefs}
\end{figure}

We use PACE to run FPCA using the irregularly and sparsely observed CVP data, yielding the mean function estimate in the left panel of Figure \ref{fig:meanCVPandFPCS}. Three FPCs of the CVP function, which explain over $90\%$ of total variance, are used to estimate the functional ZIP outcome model. The right panel of Figure \ref{fig:meanCVPandFPCS} displays the first three FPCs of the CVP function, explaining approximately $75\%$, $11\%$ and $8\%$ of the total variation, respectively. The first principal component gradually increases until around the 12-hour mark and then remains relatively flat, reflecting a general shift in the overall CVP level during the observation period. The second component captures a steady decline across time, highlighting a contrast between earlier and later phases of the trajectory. The third component shows a distinct sinusoidal pattern, decreasing in the first half before rising again, suggesting a transient dip followed by recovery in the CVP pattern.

The estimated coefficient functions from the mediator and outcome models are shown in Figure \ref{fig:estimatedcoefs}. $\widehat{\beta}_1(t)$ in the left panel represents the estimated functional coefficient for gender in the function-on-scalar regression, Model (\ref{eq:mediatormodel}), for the CVP function, which can be interpreted as the partial effect of gender on the CVP function at time $t$. An overall upward trend is evident throughout the time span, corroborating the widening gap between the mean CVP function of genders as seen in  Figure \ref{fig:readmissiondistandcvpcurveMvsF}. Recall that the functional ZIP model for 60-day rehospitalisations is of the form in (\ref{eq:outcomemodel}), (\ref{eq:logisticlink}) and (\ref{eq:loglink}), where $\gamma_1(t)$ and $\alpha_1(t)$ are the coefficient functions for the CVP function. They represent the weight of the CVP level at time $t$ in determining the rate of the Poisson component and the probability of being an excess zero in the binary component, respectively. The middle and right panels demonstrate their estimates, and a similar pattern of mild curvature can be observed.

\subsubsection{Results from causal mediation analysis}\label{subsubsec5.3.2}

\begin{table}[t]
\caption{Summary of the estimated total effect, the natural direct and indirect effects from the proposed methods implemented to the study data from MIMIC-IV: $n = 4961$, the basis system used for the mediator model is cubic splines with $nbasis = 20$, the number of functional principal components used for the outcome model $\widehat{K} = 3$, the number of simulated sets of model parameters in Algorithm \ref{alg:1} $Q = 1000$.\label{tabapplicationresults}}
\tabcolsep=0pt
\begin{tabular*}{\textwidth}{@{\extracolsep{\fill}}lccccc@{\extracolsep{\fill}}}
\toprule%
Estimand & Point Estimate  & SE & $95\%$ CI \\
\midrule
$\tau_{\textsubscript{TE}}$ & $0.153$ & $0.020$ & $(0.116, 0.195)$ \\
$\tau_{\textsubscript{NIE}}(1)$ & $0.012$ & $0.003$ & $(0.007, 0.018)$ \\
$\tau_{\textsubscript{NDE}}(0)$ & $0.141$ & $0.020$ & $(0.104, 0.182)$ \\
$\tau_{\textsubscript{NIE}}(0)$ & $0.007$ & $0.001$ & $(0.005, 0.010)$ \\
$\tau_{\textsubscript{NDE}}(1)$ & $0.146$ & $0.020$ & $(0.108, 0.188)$ \\
\midrule
\end{tabular*}
\end{table}

\begin{figure}[!t]
    \centering
    \includegraphics[width=0.95\linewidth]{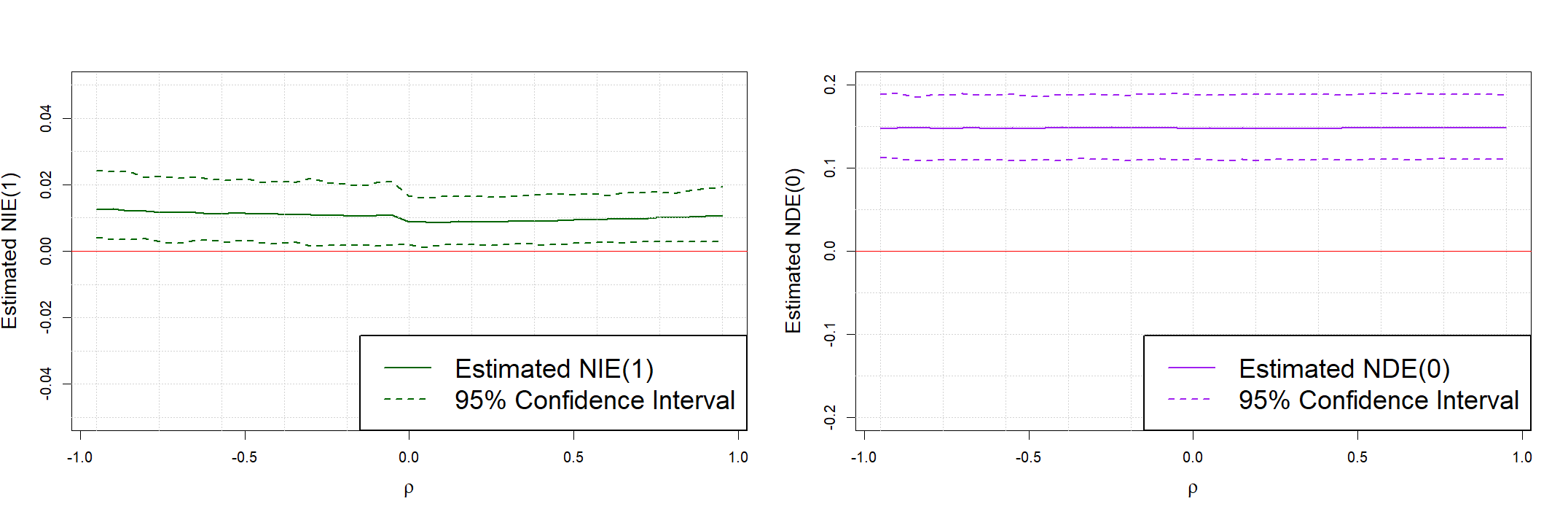}
    \caption{Estimated natural direct and indirect effects (solid line) and $95\%$ confidence intervals (dotted lines) corresponding to varying values of $\rho$ from the sensitivity analysis: $\widehat{\tau}_{\textsubscript{NIE}}^{\rho}(1)$ on the left panel; $\widehat{\tau}_{\textsubscript{NDE}}^{\rho}(0)$ on the right panel.}
    \label{fig:sensanalysis}
\end{figure}

Table \ref{tabapplicationresults} summarises the point estimate, standard error and $95\%$ confidence interval for the estimated total effect, the natural direct and indirect effects from the sparse functional causal mediation analysis with a zero-inflated count outcome. After adjusting for baseline age and race, we observe a statistically significant total effect of gender on the number of hospital readmissions within 60 days of discharge from primary CABG admission. Firstly, according to evidence found by the proposed methods, female patients, compared to male patients, in MIMIC-IV suffer an average increase of 0.153 hospital readmissions within 60 days after undergoing CABG surgery.

Secondly, the total effect is found to be significantly mediated by the smooth function underlying CVP levels measured during the first 24 hours of post-surgical ICU stay. For example, we estimate $\tau_{\textsubscript{NIE}}(1)$ to be approximately $0.012$. This means that being a female patient with a CVP level comparable to that of a man during the ICU stay after undergoing CABG surgery can reduce the number of 60-day hospital readmissions by $0.012$. Furthermore, this amounts to approximately $7.8\%$ of the total effect, indicating that $7.8\%$ of the effect of gender on 60-day rehospitalisation counts can be explained by the time-varying CVP level during the first 24 hours of the patient's post-surgical ICU stay.

In the sensitivity analysis, we examine how the estimated natural direct and indirect effects may be influenced by unmeasured confounding that induces a correlation $\rho$ between the potential mediator and potential outcome. The assumed value of $\rho$ varies from $-0.95$ to $0.95$ in increments of $0.05$, with $Q_s = 1000$ Monte Carlo iterations used in Algorithm~\ref{alg:2}. Figure~\ref{fig:sensanalysis} illustrates that the estimated causal effects in the MIMIC-IV study remain relatively stable and statistically significant across a broad range of $\rho$ values. This limited sensitivity may be partly attributable to the discrete and highly zero-inflated nature of the outcome. Even when strong dependence is introduced between the potential mediator and outcome through large values of $\rho$, the concentration of mass at zero may constrain variation in the simulated outcomes.

The above findings on the causal relationships between gender, CVP, and rehospitalisations after CABG align with existing studies on cardiac surgery, offering insights into the gender disparities in postoperative complications. It is established that women who undergo CABG surgery tend to experience more hospital readmissions, accounting for baseline age and race. We further demonstrate that this adverse effect can be partly explained by differences in CVP levels between genders during the first 24 hours of post-surgical ICU stay. These results shed light on the underlying biological factors contributing to sex disparities in postoperative outcomes, with the significant mediating effect of CVP possibly linked to differences like the thinner vessels observed in females.

\section{Discussion}\label{sec6}

To investigate the causal pathway from gender to 60-day rehospitalisation counts among patients who underwent CABG surgery in the MIMIC-IV database, we propose a causal mediation framework involving a sparse functional mediator and a zero-inflated count outcome, along with estimation procedures for the causal effects and a sensitivity analysis. Specifically, with a functional mediator, we formalise the potential outcomes framework and discuss the definition of causal effects and the identifying assumptions for the effects. A function-on-scalar regression is used to model the underlying smooth mediator process from infrequent and irregular discrete observations, and a functional zero-inflated Poisson model is used to address the previously understudied relationship between a zero-inflated count outcome and a functional mediator. Nonparametric identification results \citep{imaimed, Nguyenmedassump} are adapted to our setting with a functional mediator to facilitate the parameter-based Monte Carlo approximation algorithm for the estimation and inference of causal effects. We also develop a sensitivity analysis to assess the robustness of effect estimates to potential unmeasured confounding between the mediator and the outcome. Simulation studies validate the performance of the proposed methods in estimating the causal effects in different sample sizes and levels of complexity of the functional mediator. 

With the selected cohort of CABG patients from MIMIC-IV, we establish both a significant total effect of gender on the number of readmissions and a significant mediating effect on this pathway by the time-varying CVP level during the first 24 hours of post-surgical ICU stay. Since the early evidence from the likes of \citet{jaglal1995higher} and \citet{edwards1998impact}, it has long been acknowledged in the literature that the outcomes for women following CABG differ from those of men. In this context, our findings not only confirm the inferior outcomes in terms of rehospitalisations for female patients but also aid in the effort to uncover the underlying causes of such disparities and provide directions for bridging this gap. In particular, while numerous statistical analyses concerning the matter focus on reasons related to health equity, such as referral bias \citep{wagner2024sex}, we find that postoperative changes in CVP level, potentially as a consequence of innate biological differences between genders, also play a significant role. Consequently, evidence from our analysis suggests that clinical practitioners could monitor and manage elevated CVP levels post-surgery to reduce hospital readmissions.

It should be noted that the proposed parameter-based Monte Carlo approximation algorithm can accommodate different models for the mediator and outcome. Therefore, an immediate future direction of methodological development is to incorporate different models for the zero-inflated count outcome, such as the functional hurdle model and the functional negative binomial model. Furthermore, the algorithm may even be employed to study other types of outcomes, for example, binary outcomes, that invoke nonlinearity and call for the use of other generalised functional linear models. Another area for future research arises from an important limitation of the proposed method, which is its computational complexity. Several articles \citep{wangzero,chengzero,guo2018zero,OROURKE2019zero} have demonstrated the use of similar simulation-based approaches for causal mediation analysis with a zero-inflated count outcome in a non-functional setting, where computational intensity is already a problem. As a result of involving functional linear and nonlinear models, the use of the proposed method can be even more computationally intensive, especially for large datasets. Naturally, deriving a theoretically sound analytical solution to the causal effects with a functional mediator and a zero-inflated count outcome is an important future direction. One possible route is to adapt the marginalised ZIP model \citep{long2014marginalized} to involve a functional predictor to facilitate closed-form expressions of the causal estimands. In addition, we acknowledge the conceptual challenges associated with applying the natural effects mediation framework to non-manipulable treatments such as gender. As discussed by \citet{didden2025targeting}, interventional analogues of mediation effects require substantially weaker identifying assumptions and offer a more interpretable and policy-relevant alternative in settings with non-manipulable treatments. While the natural effects remain a valuable conceptual tool, their identification relies on stronger assumptions that may be more difficult to justify in certain contexts. Developing methodologies in settings with functional mediators and zero-inflated outcomes under the interventional paradigm represents a natural and promising avenue for future methodological development.

\bibliographystyle{abbrvnat}
\bibliography{references}

\end{document}